\documentclass{SCGE}
\usepackage{longtable}

\begin{document}

\begin{picture}(0,0){\rm
\put(0,-39){\makebox[160truemm][l]{\bf {\sanhao\raisebox{2pt}{.}}
Review  {\sanhao\raisebox{1.5pt}{.}}}}}
\put(0,-52){\jiuwuhao {\textcolor[rgb]{0.5,0.5,0.5}{\sf 
}}}
\end{picture}

\def\bm{\boldsymbol}
\def\dl{\displaystyle}
\def\du{\end{document}}
\def\pi{{\uppi}}

\Year{2012} %
\Month{December} %
\Vol{55} 
\No{12} 
\BeginPage{1} 
\EndPage{14} 
\AuthorMark{{\rm Feng W X, \textit{et al.}}}  
\AuthorMarkCite{{\rm Feng W X, \textit{et al.}}.} 
\DOI{10.1007/s11433-012-4929-9} 

\title{Three-dimensional topological insulators: A review on host materials}

\author[*]{FENG Wanxiang and YAO Yugui}{}

\address[{\rm}]{School of Physics, Beijing Institute of Technology, Beijing 100081, China}

\maketitle \vspace{-3.5mm}
{\footnotesize\begin{center} Received August 30, 2012; accepted October 22, 2012 
\end{center}}\vspace*{-5mm}

\begin{center}
\rule{16.5cm}{0.4pt}
\parbox{16.5cm}
{\begin{abstract}

In recent years, three-dimensional topological insulators (3DTI) as a novel state of quantum matter have become a hot topic in the fields of condensed matter physics and materials sciences.  To fulfill many spectacularly novel quantum phenomena predicted in 3DTI, real host materials are of crucial importance.  In this review paper, we first introduce general methods of searching for new 3DTI based on the density-functional theory.  Then, we review the recent progress on materials realization of 3DTI including simple elements, binary compounds, ternary compounds, and quaternary compounds.  In these potential host materials, some of them have already been confirmed by experiments while the others are not yet.  The 3DTI discussed here does not contain the materials with strong electron-electron correlation.  Lastly, we give a brief summary and some outlooks in further studies.

\end{abstract}}
\end{center}\vspace*{-0.6cm}

\begin{center}
\parbox{16.5cm}
{\bf\ Three dimension, topological insulator, host material}
\end{center}

\begin{center}
{}
\Cit{Feng W X. Yao Y G. Three-dimensional topological insulators: A review on host materials. Sci China-Phys Mech Astron, 2012, doi: 10.1007/s11433-012-4929-9}
\end{center}

\wuhao\vspace*{1.5mm}
\renewcommand{\baselinestretch}{1.08} \baselineskip 12.2pt\parindent=10.8pt
\renewcommand{\thefootnote}

\begin{multicols}{2}

\section{Introduction}\vspace*{-2mm}

In the past few decades, topological quantum states characterized by specially topological orders [1,2] have become one of the most critical physical phenomena in condensed matter physics.  In general concept, \textit{topological insulator} contains two classes of topological quantum states depending on whether the system has time-reversal symmetry (TRS) or not.  In the case of two-dimension, the first is the quantum Hall (QH) state [3], in which the TRS is explicit broken due to external magnetic field or internal magnetization.  This topologically nontrivial state can be classified by the first Chern number proposed by Thouless et al. [2], directly connected to the quantized Hall conductivity.  The second is the quantum spin Hall (QSH) state [4--6], which is protected by TRS and essentially resulted from spin-orbit coupling (SOC).  Similar to the role of the Chern number, Z$_2$ topological invariant in TRS protected systems was introduced by Kane and Mele [7], that is, Z$_2$=1 (0) represents topologically nontrivial (trivial) state.  The \textit{topological insulator} with the common feature,

\vspace*{0mm}
\noindent\rule{2.5cm}{0.4pt}\\[0.1mm]{\qihao *Corresponding author (YAO Yugui, email: ygyao@bit.edu.cn)}

\no having gaped insulating states in the bulk but gapless surface states on the edge, fundamentally distinguishes itself from the ordinary insulating state originally defined by Kohn [8].

The QSH state, i.e., two-dimensional topological insulator, has attracted research focus because of its promising applications in spintronic devices.  The existence of QSH state in graphene was first proposed by Kane and Mele [5,6], however, subsequent works [9--11] showed that the band-gap induced by the SOC is too small ($\sim$ 10$^{-6}$ eV) and QSH state in graphene can not be observed in current experimental conditions.  A more realistic material is the HgTe/CdTe quantum well, theoretically predicted by Bernevig et al. [12] and then experimentally confirmed by K\"{o}nig et al. [13].  Soon after the QSH state was discovered, topological quantum states under TRS were also extended in three-dimension [14-16], i.e., the three-dimensional topological insulators (3DTI).  In 3D, there are four Z$_2$ topological invariants, $\nu_{0};(\nu_{1}\nu_{2}\nu_{3})$, used to classify the topology of an insulator [17].  A nonzero $\nu_{0}$ indicates that the system is a strong topological insulator (STI).  When $\nu_{0}$=0, the systems are further classified according to $\nu_{1}$, $\nu_{2}$, and $\nu_{3}$.  The systems with $\nu_{1,2, \textrm{ or } 3}$=0 (not all of them) are called weak topological insulators (WTI), while 0;(000) is a normal insulator (NI).  A STI can not be directly connected to a WTI or a NI by any adiabatically continuous transformation of band structure.  In the WTI or NI, the surface states have an even number of Dirac points and can be violated by structural disorders or ferromagnetic impurities.  Conversely, in the STI, the surface states have an odd number of Dirac points and are topologically protected.

The 3DTI is fascinating mainly because of its exotic surface states [18--21], which are very robust and against from the extrinsic perturbations, such as structural disorders and nonmagnetic impurities.  Combining the topological surface state with superconducting state or ferromagnetic state, the 3DTI has shown great application potentials in quantum computing and spintronics fields [22--23].  So far many 3DTI host materials have been theoretically predicted and/or experimentally confirmed [see Table \ref{tab1}].  However, most of current 3DTI host materials have various disadvantages as observed in experimental observations, for example the large concentration of bulk carrier, which hinders further investigation along the direction of practical applications.  Therefore, searching for new better host materials with simpler electronic structures and simpler synthetic conditions is a key step in this field.

In this review paper, we focus on the 3DTI host materials.  We introduce general methods of searching for new 3DTI based on the density-functional theory.  We also discuss many theoretically proposed 3DTI materials including simple elements, binary compounds, pseudo-binary compounds, ternary compounds, and quaternary compounds.  In these potential host materials, some of them have already been confirmed by experiment while the others are not yet.  The 3DTI discussed here does not contain the materials with strong electron-electron correlation.

\section{How to search for 3D-topological insulators\label{sec:2}}

As a first step towards various applications of 3DTI, material realization is of crucial importance.  During the search of 3DTI in real materials, first-principles calculations guided by the topological band theory [18,19] have played a pivotal role.  We briefly introduce four widely used methods to search 3DTI.  In practice, two or more among these methods are usually combined together to determine the band topology of a crystal.

\subsection{Surface state electronic structure\label{sec:2.1}}

As a fundamental feature in 3DTI, there is gapless surface state inside the bulk band-gap [18--21].  The gapless surface state has the form of linear Dirac cone, which holds massless relativistic particles described by the Dirac equation [24,25].  Furthermore, the spin and momentum of surface Dirac electrons are locked together in the sense that the electrons with opposite spin must propagate in opposite direction along the boundary due to TRS.  Hence, one can determine whether an insulator is a 3DTI or not by calculating the surface state electronic structure.  An odd number of surface states crossing the Fermi level indicates a 3DTI, otherwise it is a normal insulator.  At the experimental aspect, the surface state electronic structure can be directly measured by the angle-resolved photoemission spectroscopy (ARPES).  The comparison between theoretical calculations and experimental ARPES measurements becomes possible.

Although the bulk-boundary correspondence as a basic routine can be used to find the topologically nontrivial materials, it is not effective in first-principles calculations.  Physically speaking, the surface band dispersion is sensitive to the details of the surface, such as the terminations and orientations [26--28].  Furthermore, some topologically trivial states, for example the dangling bonds on the surface of group-IV or III-V semiconductors, may manifest themselves in the surface band structure.  The surface reconstruction or hydrogen passivation can partly destroy these dangling bonds, but the coexistence of topologically trivial and nontrivial surface states certainly complicates the identification of the topological order.  From the computational point of view, first-principles calculation of surface state electronic structure is an inefficient method because it generally requires a large amount of computational resources.

\subsection{Adiabatic continuity band transformation\label{sec:2.2}}

The argument of adiabatic continuity was once used to identify the quantum spin Hall phase in HgTe/CdTe quantum well [12] and silicene [29], but can also be introduced into the 3DTI.  The basic idea is that the Hamiltonian of a system can be adiabatically connected to that of another system.  For the former and latter systems, the topological invariants can only change when the bulk band-gap closed.  Therefore, if the band structure of a material can be smoothly transformed to that of another one without closing the band-gap, then they must share the same topological classification.  In contrast, if the band-gap closes during the band transformation, the topological phase transition may occur.

In practice, the adiabatic continuity transformation of band structure can be artificially controlled, for example, by altering the composition $x$ in solid alloy Bi$_{1-x}$Sb$_{x}$ [17,30,31] or Pb$_{1-x}$Sn$_{x}$Te [17].  By starting $x$ from zero, these systems undergo the topological phase transition from trivial state to nontrivial state.  If we continue increasing $x$ for Pb$_{1-x}$Sn$_{x}$Te, a further transition from nontrivial state to trivial state will happen.  Tuning other parameters, such as atomic number and strength of SOC, are also widely used in literatures.  The obvious advantage of this method is its apparent ease.  One can predict new 3DTI candidates based on some well-known topologically nontrivial or trivial materials.  While many intermediate phases are required along the transformational path connecting the initial state and the final one, which makes this method computationally demanding.

\subsection{Band inversion picture\label{sec:2.3}}

Sometimes, topologically nontrivial materials can be well recognized at the first glance by the so-called band inversion picture [12].  In most of the common semiconductors, the valence band-edges are formed by the $p$-orbits of electrons, whereas the conduction band-edges are formed by the $s$-orbits of electrons.  This situation belongs to normal band order, i.e., topologically trivial phase.  In other cases, the relativistic effect from heavy elements can be so large that the $s$-orbits is pushed below the $p$-orbits, that is, the inverted band order appears.  A typical example is HgTe in which the effective positive charge of Hg core is increased due to the partly delocalization of its $d$-orbits.  This causes the $s$-orbits to be more attracted by the Hg core, and consequently, the $s$-orbits are pushed below the $p$-orbits. i.e., the inverted band structure forms [32].

The band inversion is a strong indication that a material falls into topologically nontrivial phase.  The first application is the prediction of quantum spin Hall effect in HgTe/CdTe quantum well proposed by Bernevig et al. [12].  The quantum well structure is that the HgTe with inverted band order is sandwiched by the CdTe with normal band order.  The topological property of entire quantum well is determined by the thickness $d$ of HgTe layer.  The critical thickness $d_c$ is predicted to be about 6.5 nm [12].  For a thin well when $d<d_c$, the CdTe has the dominant effect and the entire quantum well is topologically trivial with normal band order, while for a thick well when $d>d_c$, the HgTe has a critical role role and the entire quantum well is topologically nontrivial with inverted band order.

Similar to HgTe and CdTe, the band topology of other cubic semiconductors with zinc-blende-like structure can also be determined by the band inversion picture.  To do this, one need to define a useful physical quantity, namely, the band inversion strength (BIS).  Specially at $\Gamma$ point, the BIS is defined as the energy differences between $\Gamma_6$ state (formed by $s$-orbits) and $\Gamma_8$ state (formed by $p$-orbits) [33,34] i.e.,
\begin{equation}
\Delta E=E_{\Gamma_{6}}-E_{\Gamma_{8}}\;.
\end{equation} In general, negative $\Delta E$ typically indicates that the materials are in topologically nontrivial phase, while those with positive $\Delta E$ are in topologically trivial phase.

Moreover, the conventional $sp$-orbits semiconductors, the band inversion has also been found in $pp$-orbits and $df$-orbits semiconductors.  In Bi$_2$Se$_3$ family, both the valence and conduction band-edges are formed by $p$-orbits but with opposite parities.  It was predicted that Bi$_2$Se$_3$ family is 3DTI due to the inverted band order between $\left|P1_{z}^{+}\right\rangle $ and $\left|P2_{z}^{-}\right\rangle $ orbits [35].  Here, the $+$ ($-$) means the parity of the corresponding Bloch wavefuntions.  The band inversion can be interpreted that at $\Gamma$ point these two $p$-orbits with opposite parities are exchanged when SOC turns on.  Other examples are actinide compounds Am$X$ ($X$=N, P, As, Sb, and Bi) and Pu$Y$ ($Y$=Se and Te), which were recently discovered by Zhang et al. [36] as a new class of 3DTI driven by strong electron-electron interactions.  In these compounds, the $\Gamma_{8}^{+}$ state of 6$d$-orbit locates below the $\Gamma_{8}^{-}$ state of 5$f$-orbit at $X$ point, and as a result, the inverted band order occurs.

Although the band inversion at some high-symmetry points is a convenient way to find topologically nontrivial materials, for example HgTe/CdTe quantum well, Bi$_2$Se$_3$ family, and actinide compounds, it should be used with more care because the band topology is a global property within the entire Brillouin zone and is not only limited to some high-symmetry points.

\subsection{Z$_2$ topological invariants\label{sec:2.4}}

The most general and direct method of searching for 3DTI is to calculate the $Z_{2}$ topological invariants from the bulk band structure [14--16].  There exist two situations depending on whether the system has spatial inversion symmetry or not.  If a system has inversion symmetry, the calculation of $Z_{2}$ topological invariants can be well simplified based on the parity criterion developed by Fu and Kane [17].  Conversely, if a system does not have inversion symmetry, Fukui and Hatsugai [37] have developed an effective algorithm, which requires Bloch functions (BFs) on a dense two-dimensional grid to compute the Z$_{2}$ topological invariants.  The implementation of these two methods within full-potential linearized augmented plane-wave (FP-LAPW) formalism has been demonstrated [38].  In the following, we briefly introduce these two methods.

In the systems with inversion symmetry, the parity analysis only requires the knowledge of BFs on the time-reversal invariant momenta (TRIM) in the Brillouin zone (BZ) [17].  In 3D system, there are eight TRIMs, $\mathbf{\Gamma}_{i=\left(n_{1}n_{2}n_{3}\right)}=\frac{1}{2}\left(n_{1}\bm{G}_{1}+n_{2}\bm{G}_{2}+n_{3}\bm{G}_{3}\right)$, where $\bm{G}_{j}$ are primitive reciprocal-lattice vectors with $n_{j}=0\textrm{ or }1$.  The $Z_{2}$ invariants are determined by the quantities

\begin{equation}
\delta_{i}=\prod_{m=1}^{N_{occ}}\xi_{2m}\left(\Gamma_{i}\right).\label{eq:delta_i}
\end{equation} Here, $\xi_{2m}\left(\Gamma_{i}\right)=\left\langle \Psi_{2m,\Gamma_{i}}\left|P\right|\Psi_{2m,\Gamma_{i}}\right\rangle $ is the eigenvalue of parity operator $P$ at the \textit{2m}-th occupied band and TRIMs $\Gamma_{i}$.  The $\xi_{2m}\left(\Gamma_{i}\right)$ is equal to 1 (-1), corresponding to even (odd) parity of the BFs.  The sum is over all of the occupied bands with only even band index due to the Kramers degeneracy at TRIMs.  In 3D system, there are four independent invariants $\nu_{0};(\nu_{1}\nu_{2}\nu_{3})$, given by [17]

\begin{equation}
\left(-1\right)^{\nu_{0}}=\prod_{i=1}^{8}\delta_{i},\label{eq:Z2_v0}
\end{equation}

\begin{equation}
\left(-1\right)^{\nu_{k}}=\prod_{n_{k}=1,n_{j\neq k}=0,1}\delta_{i=\left(n_{1}n_{2}n_{3}\right)},\label{eq:Z2_vk}
\end{equation} where $\nu_{0}$ is independent of the choice of primitive reciprocal-lattice vectors $\bm{G}_{j}$ while $\nu_{1}$, $\nu_{2}$, and $\nu_{3}$ are not.  The combination of these four independent invariants $\nu_{0};(\nu_{1}\nu_{2}\nu_{3})$ clearly distinguish three classes of states: STI, WTI, and NI [17].

\begin{figure}[H]
\centering
\includegraphics[width=0.9\columnwidth]{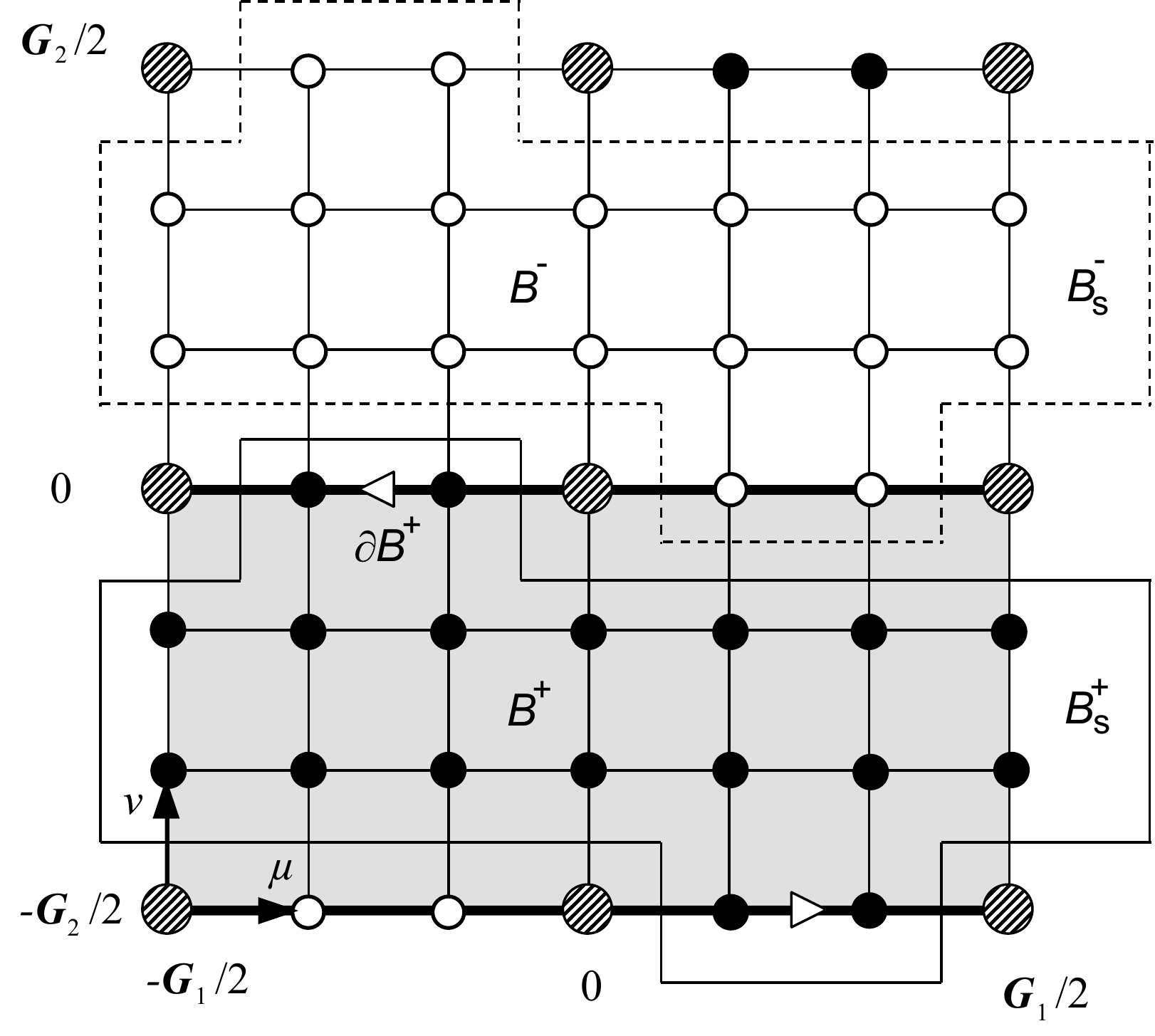}
\caption{Schematic drawing of lattice mesh in a 2D-torus.  Under the time-reversal constraint, only half of 2D-torus $\mathcal{B}^{+}$ is needed, which is denoted by shaded region.  The thick lines indicate the boundary of $\mathcal{B}^{+}$, i.e., $\partial\mathcal{B}^{+}$, and the open arrows denote their directions.  All $\bm{k}$-points are divided into three classes: $\mathcal{B}_{s}^{+}$, $\mathcal{B}_{s}^{-}$, and $\mathcal{B}_{s}^{0}$, which are represented by small solid, small open and large shaded circles, respectively.  Adapted from Ref. [38].}\label{fig1}
\end{figure}
In the systems without inversion symmetry, an effective algorithm on a dense two-dimensional grid to compute the Z$_{2}$ invariants has been proposed by Fukui and Hatsugai within a tight-binding framework [37].  It was shown that under the time-reversal constraint the $Z_{2}$ invariants can be written in terms of the Berry gauge potential and Berry curvature associated with the BFs [39],

\begin{equation}
Z_{2}=\frac{1}{2\pi}\left[\oint_{\partial\mathcal{B}^{+}}d\bm{k}\cdot\bm{\mathcal{A}}\left(\bm{k}\right)-\int_{\mathcal{B}^{+}}d^{2}k\,\mathcal{F}\left(\bm{k}\right)\right]\textrm{ mod 2},\label{eq:Z2}
\end{equation}  where $\bm{\mathcal{A}}\left(\bm{k}\right)$ and $\mathcal{F}\left(\bm{k}\right)$ are the Berry connection and Berry curvature, respectively,

\begin{equation}
\bm{\mathcal{A}}\left(\bm{k}\right)=i\sum_{n}\left\langle u_{n}\left(\bm{k}\right)\mid\bm{\nabla}_{\bm{k}}u_{n}\left(\bm{k}\right)\right\rangle \label{eq:Berry_conn}
\end{equation} and

\begin{equation}
\mathcal{F}\left(\bm{k}\right)=\bm{\nabla}_{\bm{k}}\times\bm{\mathcal{A}}\left(\bm{k}\right)\mid_{z}.\label{eq:Berry_curv}
\end{equation} The $\mathcal{B}^{+}$ and $\partial\mathcal{B}^{+}$ indicate half of 2D-torus and its boundary, respectively (see Fig. \ref{fig1}).  By using the periodic gauge [40,41]

\begin{equation}
\left|u_{n}\left(\bm{k}+\bm{G_{j}}\right)\right\rangle =e^{-i\bm{G_{j}\cdot r}}\left|u_{n}\left(\bm{k}\right)\right\rangle .\label{eq:Peri_gauge}
\end{equation} and two time-reversal constraints [37,39]

\begin{equation}
\begin{array}{cc}
\left|u_{n}\left(-\bm{k}\right)\right\rangle =\Theta\left|u_{n}\left(\bm{k}\right)\right\rangle , & \bm{k}\in\mathcal{B}_{s}^{+}\end{array},\label{eq:Time_con1}
\end{equation} and

\begin{equation}
\begin{array}{cc}
\left|u_{2n}\left(-\bm{k}\right)\right\rangle =\Theta\left|u_{2n-1}\left(\bm{k}\right)\right\rangle , & -\bm{k}\textrm{ and }\bm{k}\in\mathcal{B}_{s}^{0}\end{array},\label{eq:Time_con2}
\end{equation} one can obtain the periodic part of BFs $\left|u_{n}\left(\bm{k}\right)\right\rangle $ at every $\bm{k}$-point of the 2D-torus.  Then, the Berry connection [Eq. (\ref{eq:Berry_conn})] and Berry curvature [Eq. (\ref{eq:Berry_curv})] can be calculated by the finite element expressions [37].  After that, the Z$_2$ invariants can be obtained by inserting Eqs. (\ref{eq:Berry_conn}) and (\ref{eq:Berry_curv}) into Eq. (\ref{eq:Z2}).  In 3D system, there are six 2D-tori $T(Z_{0})$, $T(Z_{1})$, $T(X_{0})$, $T(X_{1})$, $T(Y_{0})$, and $T(Y_{1})$, supporting six $Z_{2}$ invariants $z_{0}$, $z_{1}$, $x_{0}$, $x_{1}$, $y_{0}$, and $y_{1}$.  However, out of these six possible Z$_{2}$ invariant only four of them are independent due to the constraint $x_{0}$ + $x_{1}$ = $y_{0}$ + $y_{1}$ = $z_{0}$ + $z_{1}$ (mod 2).  The $Z_{2}$ invariants are denoted in another way with $\nu_{0}=(z_{0}+z_{1})\textrm{ mod 2}$, $\nu_{1}=x_{1}$, $\nu_{2}=y_{1}$ and $\nu_{3}=z_{1}$, i.e., $\nu_{0};(\nu_{1}\nu_{2}\nu_{3})$ [14--16].  By using topological invariants $\nu_{0};(\nu_{1}\nu_{2}\nu_{3})$, one can determine the band topology for a given material.

Comparing with the surface state electronic structure (Sec.~\ref{sec:2.1}), adiabatic continuity band transformation (Sec.~\ref{sec:2.2}), and band inversion picture (Sec.~\ref{sec:2.3}), the calculation of the Z$_2$ topological invariants is a direct evidence for determining topologically nontrivial state.  Furthermore, from the viewpoint of first-principles calculation, it is the most computationally efficient.  Our implementation of the calculation of Z$_2$ topological invariants within FP-LAPW formalism has been recently applied to predict ternary half-Heusler [33,34] and chalcopyrite [42] 3DTI and quantum spin Hall effect in silicene thinfilm [29].  Since the critical factor is to calculate the eigenvalues of the parity operator and time-reversal operator in the systems with and without inversion symmetry respectively, it can be implemented in any other first-principles methods, such as the pseudopotential planewave and the linear muffin-tin orbital.  There appears another approach to compute the $Z_{2}$ invariants by employing the charge center of Wannier functions [43,44].

\section{3DTI host materials\label{sec:3}}

We review the 3DTI host materials in the sequence of simple elements, binary compounds, pseudo-binary compounds, ternary compounds, and quaternary compounds.  In each part, only one or several typical materials are discussed.  Table \ref{tab1} includes a list of all 3DTI host materials.  In these potential host materials, some of them have been confirmed by experiments while the others are not yet.  The readers can find the interesting materials in Table \ref{tab1} and refer to the literatures therein.  It also should be noted that the 3DTI discussed here does not contain the materials with strong electron-electron correlation.

\subsection{3DTI in simple elements\label{sec:3.1}}

\subsubsection{$\alpha$-Sn}

\begin{figure}[H]
\centering
\includegraphics[width=0.9\columnwidth]{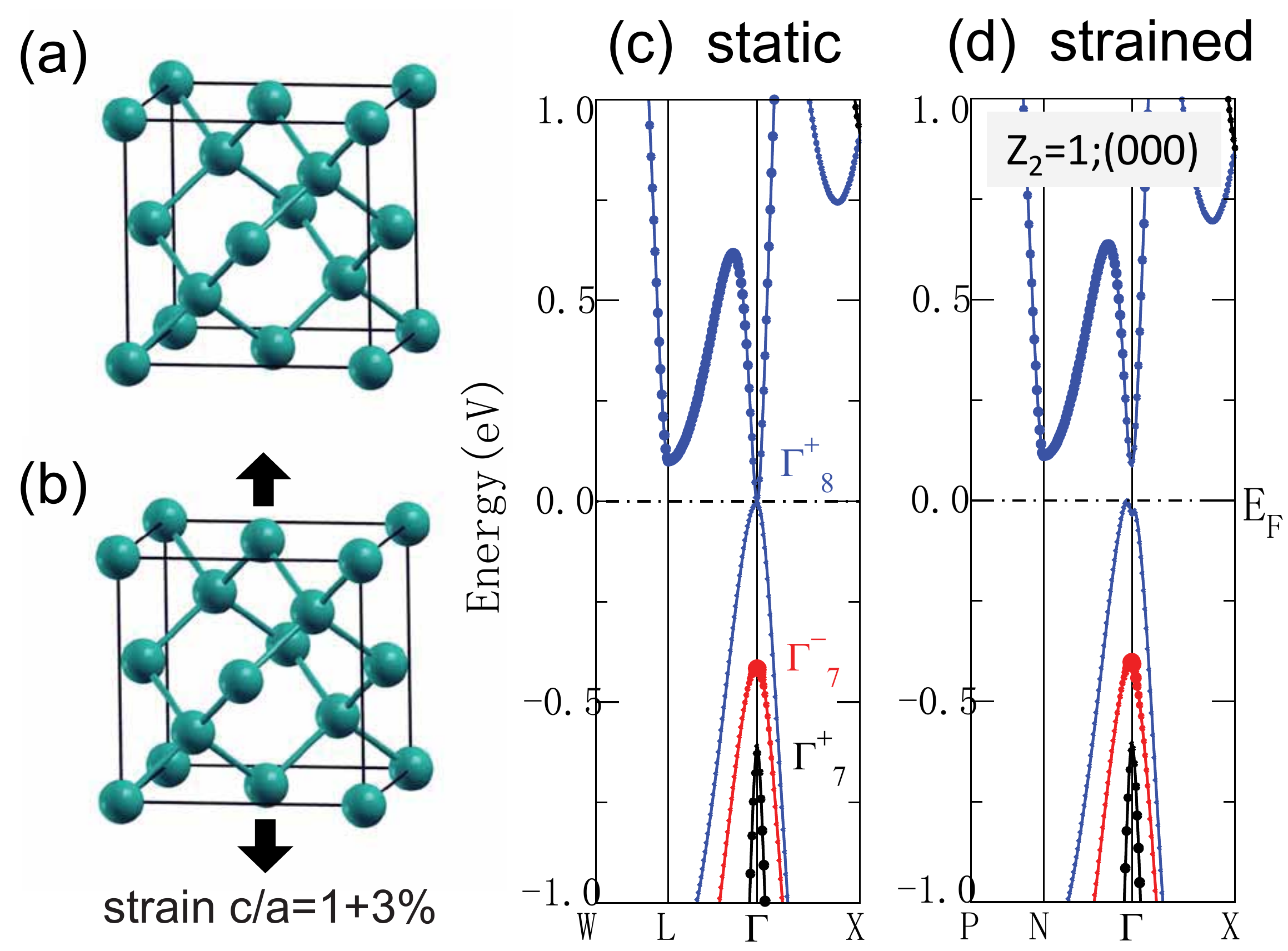}
\caption{Crystal structure of $\alpha$-Sn at its static lattice constant (a) and under a uniaxial strain c/a=1+3\% (b), respectively.  (c) and (d) are the corresponding band structures with the $s$-orbits projections, i.e., the size of solid circles.  The calculated Z$_2$ topological invariant 1;(000) indicates a strong 3DTI.  Here, the full-relativistic band structures were performed using FP-LAPW method [45] with the MBJLDA exchange-correlation potential [46], implemented in \textsc{wien2k} package [47].}\label{fig2}
\end{figure}

The group-IV element $\alpha$-Sn (grey tin) crystallizes in the diamond structure with space group Fd$\bar{3}$m (No. 227), as shown in Fig. \ref{fig2}(a).  The full-relativistic band structure in Fig. \ref{fig2}(c) clearly shows that at the static lattice constant $\alpha$-Sn is a zero band-gap semiconductor with the feature of band inversion.  The valence band maximum (VBM) and the conduction band minimum (CBM) are degenerated at $\Gamma$ point and the Fermi level crosses the fourfold-degenerated $\Gamma_{8}^{+}$ state (from $p$-orbits).  The $s$-orbits-like $\Gamma_{7}^{-}$ state is situated below the $\Gamma_{8}^{+}$ state, forming the inverted band structure [48,49].  Hence, the band structure of $\alpha$-Sn is qualitatively different from those of the other group IV elements C, Si, and Ge [49].

Fu et al. [17] proposed that a uniaxial strain can tune $\alpha$-Sn into a 3DTI.  The basic idea is that after applying a uniaxial strain the fourfold-degenerated $\Gamma_{8}^{+}$ state is lifted and a band-gap opens around $\Gamma$ point. Consequently, the topologically nontrivial state forms because the band inversion does not change.  Fig. \ref{fig2}(d) shows the band structure of $\alpha$-Sn under a uniaxial strain c/a=1+3\% but with constant volume.  One can see that the band-gap opens while the band inversion preserves.  To further confirm the argument of band inversion, the Z$_2$ topological invariant is calculated.  The result 1;(000) indicates that the strained $\alpha$-Sn is indeed a STI.  Although the theoretical picture is very clear, there is not any experimental work on $\alpha$-Sn currently.

\subsubsection{Sb}

\begin{figure}[H]
\centering
\includegraphics[width=0.9\columnwidth]{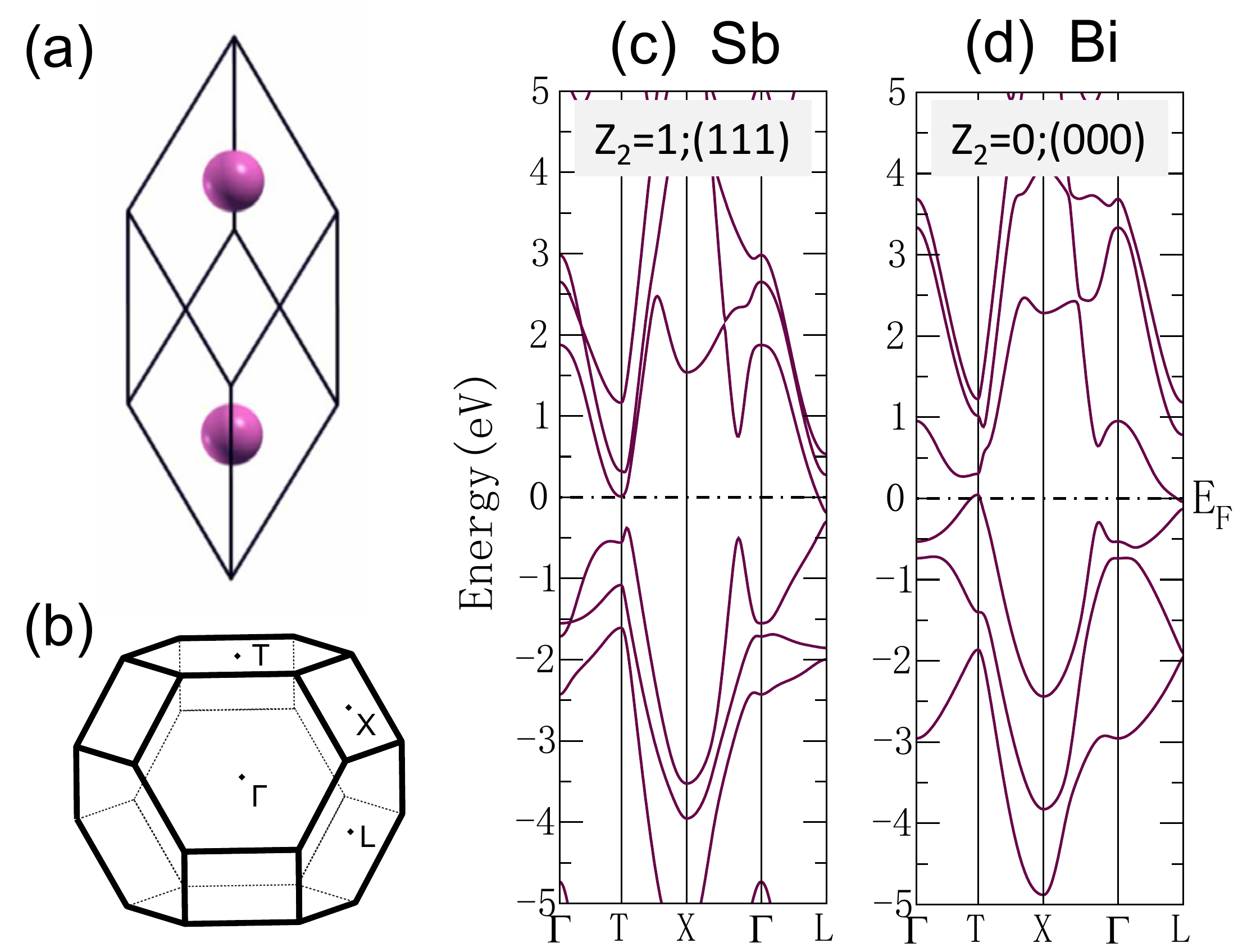}
\caption{Crystal structure (a) and the Brillouin zone (b) of Sb (Bi).  (c) and (d) are the band structures of Sb and Bi, respectively.  The Sb is a STI with Z$_2$ topological invariant 1;(111), while Bi is a NI with 0;(000).  Here, the full-relativistic band structures were performed using FP-LAPW method [45] with the PBE-GGA exchange-correlation potential [50], implemented in \textsc{wien2k} package [47].}\label{fig3}
\end{figure}

The Sb (antimony) crystallizes in rhombohedral structure with space group R$\bar{3}$m (No. 166), as shown in Fig. \ref{fig3}(a).  The full-relativistic band structure in Fig. \ref{fig3}(c) shows Sb is a semimetals with small pockets of electron in the vicinity of the L point.  There are also small pockets of hole around the low-symmetry H point (not shown).  Although it is not an insulator, Z$_2$ topological invariant can still be defined because the local band-gaps (negative indirect band-gap) exist at every $\bm{k}$-point throughout the entire BZ [17].  The result 1;(111) indicates that the Sb is a STI.  As another group-V element, Bi (bismuth) has similar crystal and band structure comparing to Sb, as shown in Fig. \ref{fig3}(a) and \ref{fig3}(d) respectively, but Bi is a NI with Z$_2$ topological invariant 0;(000).  The different topological classes between Sb and Bi were well interpreted by their different parities at L points [17].

Hsieh et al. [51,52] have measured the bulk and surface electronic structure of Sb.  Because bulk Sb is a semimetal, it is difficult to separate its surface state from the projection of bulk state around the Fermi level.  While by using advanced spin-resolved ARPES technique, Hsieh et al. [51] successfully isolated the surface states of Sb(111) from its bulk states over the entire BZ.  They also directly found the gapless and spin-splitting surface state, which characterizes the topologically nontrivial feature in Sb [51].

\subsection{3DTI in binary compounds\label{sec:3.2}}

\subsubsection{Bi$_{1-x}$Sb$_{x}$ alloy}

Both Sb and Bi are group-V semimetals with the rhombohedral crystal structures and similar lattice constants, therefore, they can easily form solid alloy Bi$_{1-x}$Sb$_{x}$ [53].  By altering the concentration $x$ of Sb substitution, Bi$_{1-x}$Sb$_{x}$ undergoes a phase transition between semimetal and semiconductor.  In the range of 0.07 $< x <$ 0.22, Bi$_{1-x}$Sb$_{x}$ becomes semiconductor with the largest global band-gap of 0.03 eV at $x$=0.18.  Based on a tight-binding model [54], Fu et al. [17,30] predicted that Bi$_{1-x}$Sb$_{x}$ is the first realistic 3DTI material.  Subsequently, this prediction was confirmed by first-principles calculation [31] and experimental observations [55--57].

Although both theories and experiments got a consistent conclusion that the topologically nontrivial phase exists in Bi$_{1-x}$Sb$_{x}$ alloy, the discrepancies about surface electronic structure still remain.  The main concern is the surface band configuration, including the numbers of surface bands and the crossing times between the surface bands and the Fermi level along the $\bar{\Gamma}$-$\bar{M}$ line in surface BZ.  All the experimental works [55--57] reported three surface bands ($\Sigma_{1}$, $\Sigma_{2}$, and $\Sigma_{1}^{\prime}$) lying inside band-gap of bulk projection, but with different crossing times with the Fermi level, five times in Refs. [55,56] and three times in Ref. [57].  Conversely, both tight-binding [30] and first-principles [31] calculations can not reproduce the presence of the third surface band $\Sigma_{1}^{\prime}$.  Teo et al. [30] and Zhang et al. [31] give three and five times of band crossing, respectively.  Therefore, more experimental measurements and theoretical calculations are needed for revealing the microscopic physics in Bi$_{1-x}$Sb$_{x}$.

\begin{figure}[H]
\centering
\includegraphics[width=0.9\columnwidth]{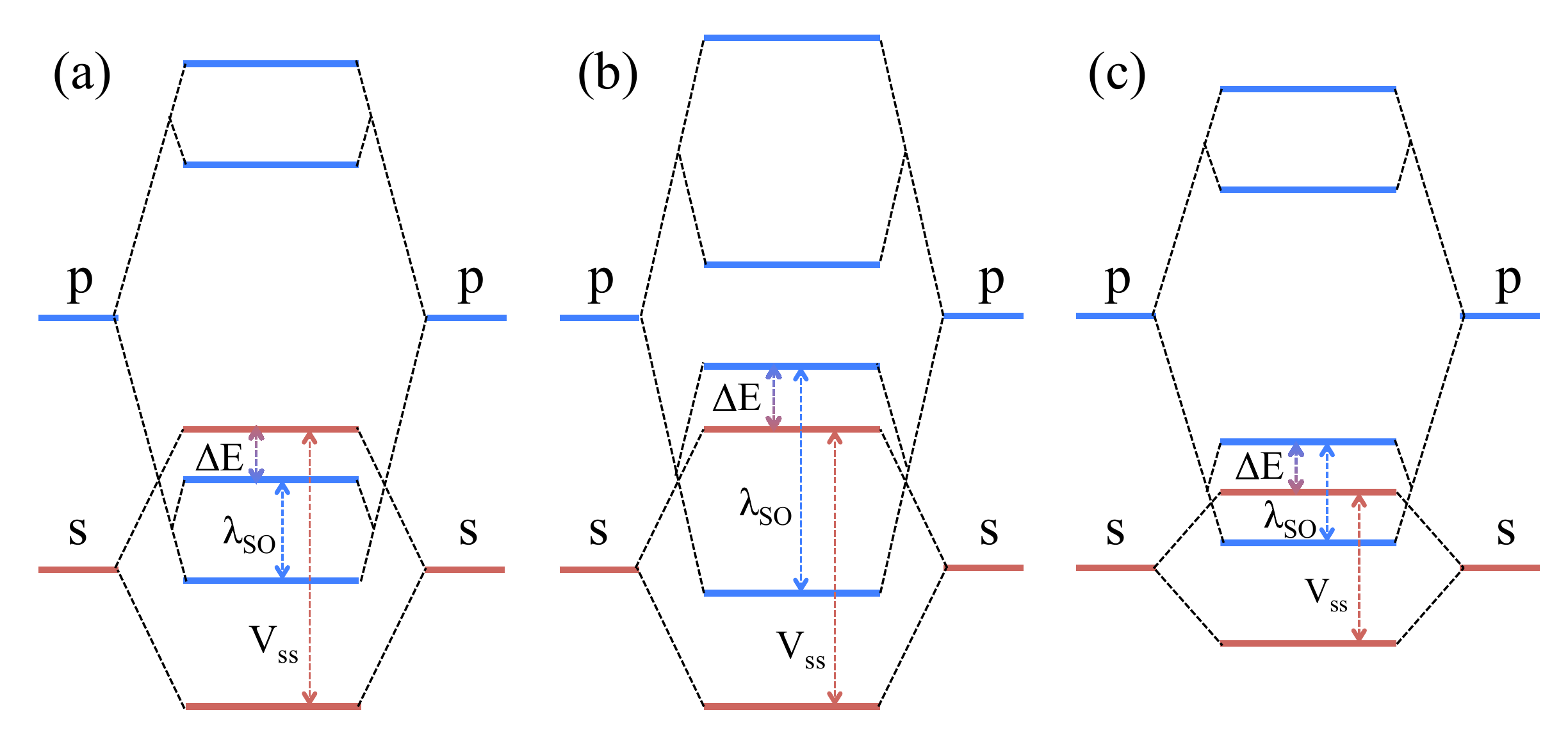}
\caption{The band order of cubic diamond/zinc-blende semiconductors at the $\Gamma$ point.  (a) The normal band order.  (b)-(c) The inverted band order can be obtained from the normal band order by either increasing the strength of SOC (b) or by decreasing the coupling potentials of $ss$-orbits or $pp$-orbits (c).  The band inversion strength is defined as $\Delta E=E_{\Gamma_6}-E_{\Gamma_8}$, which has positive value in normal band order and negative value in inverted band order.  Adapted from Ref. [75].}\label{fig4}
\end{figure}

\subsubsection{Bi$_2$Se$_3$ family}

Since the surface electronic structure of Bi$_{1-x}$Sb$_{x}$ is rather complicated and its bulk band-gap is too small, searching for new 3DTI host materials with simple surface electronic structure and large band-gap becomes extremely important.  Fortunately, Bi$_2$Se$_3$ family compounds with larger band-gap and simper surface spectrum have been found as a second generation of 3DTI materials [35,58--60], which support further study on various topologically protected phenomena at room temperature.

Tetradymite semiconductors Bi$_2$Se$_3$, Bi$_2$Te$_3$, and Sb$_2$Te$_3$, have a rhombohedral crystal structure with space group R$\bar{3}$m (No. 166).  Although Sb$_2$Se$_3$ essentially has an orthorhombic crystal structure, a virtual rhombohedral crystal structure is used at here for comparing its topological property with tetradymite semiconductors.  Zhang et al. [35] theoretically proposed that Bi$_2$Se$_3$, Bi$_2$Te$_3$, and Sb$_2$Te$_3$ are 3DTI while Sb$_2$Se$_3$ is not.  More importantly, the band-gap of Bi$_2$Se$_3$ is as large as 0.3 eV, which is larger than the energy scale of room temperature.  Almost at the same time, these family compounds were experimentally confirmed with the observation of single Dirac cone on the surface for Bi$_2$Se$_3$ [58], Bi$_2$Te$_3$ [59,60], and Sb$_2$Te$_3$ [60].  Because of the large band-gap and simple surface electronic structure, Bi$_2$Se$_3$ family compounds have been extensively studied ranging from bulk to thinfilm [61--71].

\subsubsection{HgTe and strained InSb}

HgTe is a zinc-blende compound with space group F$\bar{4}$3m (No. 216).  It is a zero band-gap semiconductor due to large relativistic effect.  The twofold-degenerated $\Gamma_{6}$ state is located below the fourfold-degenerated $\Gamma_{8}$ state, forming the inverted band order [see Fig. \ref{fig4}(b)].  Fu et al. [17] theoretically predicted that HgTe is a 3DTI under a uniaxial strain.  The basic routine is similar to that of $\alpha$-Sn, that is, the uniaxial strain is used to break the cubic symmetry and open a band-gap at $\Gamma$ point without changing the band inversion.  Recently, Br\"{u}ne et al. [72] successfully observed the Dirac-like topological surface state on strained HgTe by ARPES measurement.

InSb is also a zinc-blende semiconductor, but in contrast to HgTe, it has a small band-gap of 0.235 eV [74].  The $\Gamma_{6}$ state is located above the $\Gamma_{8}$ state, forming the normal band order [see Fig. \ref{fig4}(a)].  Feng et al. [75] proposed that InSb can be turned into a 3DTI by a 2\% $\sim$ 3\% biaxial lattice expansion.  The generic guiding principle is that lattice expansion decreases the coupling potentials of $ss$-orbits or $pp$-orbits, which leads to the band inversion [see Fig. \ref{fig4}(c)].  Here, we use different modes of strain in HgTe and InSb.  In the former one, the uniaxial strain is just used to break the cubic symmetry.  While in the latter one, the nonhydrostatic (2\% $\sim$ 3\% biaxial) lattice expansion is twofold because it not only changes normal band order to inverted band order but also breaks the cubic symmetry.

\subsection{3DTI in pseudo-binary compounds\label{sec:3.3}}

Although Bi$_2$Se$_3$ family as the second generation of 3DTI has attracted research focus because of its simple surface electronic structure, it hard to be applied in various real devices.  The prominent difficulty is high bulk carrier concentration, which remarkably masks the contribution of surface carriers in the surface transport measurement [55,62,80--82].  The high bulk carriers concentration mainly originates from the native crystal defects [58--60].  Some experimental techniques including compensate doping in bulk [59,63] and tuning Fermi level by gate voltage [83--85] have been tested, but this issue is not perfectly solved.

Recently, a promising platform for reducing the bulk carrier concentration, Bi$_{2}$Te$_{2}$Se, has been proposed by Ren et al. [86] and Xiong et al. [87].  Bi$_{2}$Te$_{2}$Se has a large bulk resistivity because its chemical characteristic makes the formation of crystal defects more difficult.  Motivated by this observation, there appears other tetradymite-like $M_{2}X_{2}Y$ ($M$ = Bi or Sb; $X$ and $Y$ = S, Se or Te) compounds [88--90] and Bi$_{2-x}$Sb$_{x}$Te$_{3-y}$Se$_{y}$ alloy [91--96].  The alloy system with a series of special combinations of $x$ and $y$ can further improve the bulk resistivity, and thus provide a diverse platform for investigating the surface transport phenomena.

It should be noted that in this review paper these tetradymite-like materials are classified into pseudo-binary 3DTI because their crystal and electronic structures are very similar to that of binary Bi$_2$Se$_3$ family.

\subsection{3DTI in ternary compounds\label{sec:3.4}}

\subsubsection{Half-Heusler}

\begin{figure*}
\includegraphics[width=2\columnwidth]{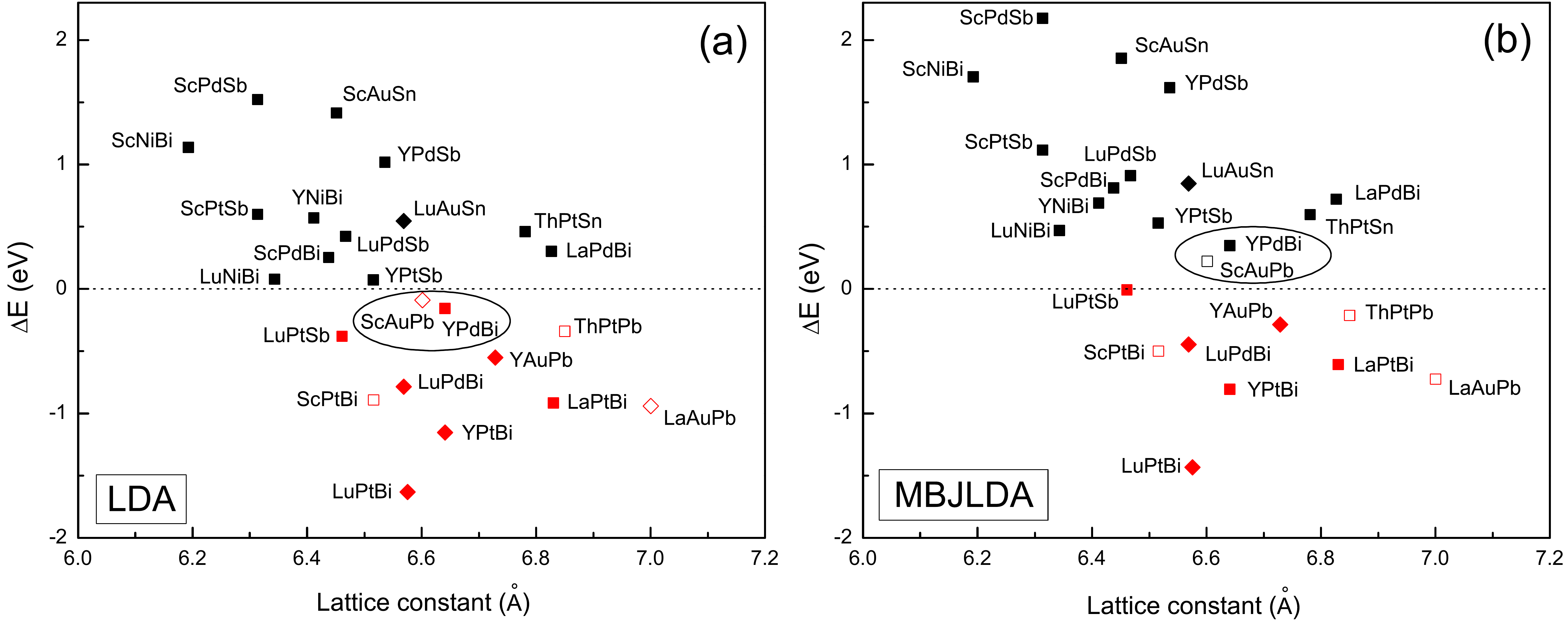}
\caption{Band inversion strength of half-Heusler compounds calculated by LDA (a) and MBJLDA (b). The materials with open symbols indicate that there is not experimental reports, and the lattice constant are obtained by first-principles total energy minimization. The red squares mark the topological insulator candidates (after applying a proper uniaxial strain), red diamonds mark topological metals, black squares mark ordinary insulators, and black diamonds mark ordinary metals. The topologically nontrivial phase is below the horizontal line, while the trivial phase is above horizontal line. The materials enclosed by circle indicate that their topological phase predicted change by using different exchange-correlation potential. Adapted from Ref. [34].}\label{fig5}
\end{figure*}

The crystal structure of ternary half-Heusler compounds is described by space group $F\bar{4}3m$ (No. 216).  The chemical formula of these materials is $XYZ$, where $X$ and $Y$ are transition or rare earth metals and $Z$ a heavy element.  It can be regarded as a hybrid compound of $XZ$ with rock-salt structure, and $XY$ and $YZ$ with the zinc-blende structure.  The band structure of half-Heusler compounds at the $\Gamma$ point near the Fermi level splits into twofold-degenerated $\Gamma_{6}$, twofold-degenerated $\Gamma_{7}$, and fourfold-degenerated $\Gamma_{8}$ states.  Away from $\Gamma$ point, the valence bands and conduction bands are well separated throughout entire BZ.

Three research groups independently predicted that under a uniaxial strain ternary half-Heusler compounds are new family of 3DTI [33,97,98].  Since the low-energy electronic structure of half-Heusler compounds is dominated at $\Gamma$ point and similar to other zinc-blende semiconductors.  The identification of topologically nontrivial feature via the band inversion picture becomes possible [97,98], just as what occurs in HgTe.  The authors in Ref. [33] also directly calculated the Z$_2$ topological invariants from the bulk band structure.  The calculated Z$_2$ topological invariant 1;(000) definitively confirmed that some of ternary half-Heusler compounds are STI.

The band topology calculated by first-principles method is sensitive to the exchange-correlation potential [34,99]. Hence, different potentials should be used to explore the topological nature.  The recently developed semilocal MBDLDA potential is believed to be better suited for calculating the topological band structure [34,99].  As shown in Fig. \ref{fig5}, the topologically nontrivial and trivial phases are located below and above the horizontal line, respectively.  One can see that the topological phases of ScAuPb and YPdBi change by using different exchange-correlation potentials.

Recently, there have appeared a few experimental works about the electronic structures, transport properties, and topological phenomena of ternary half-Heusler compounds [100--104].  Since a large number of materials with half-Heusler structure possess additional properties such as magnetism [105] and superconductivity [106], the combination of the predicted topological order with ferromagnetic order and/or superconductive order may provide an exciting platform for novel quantum devices.

\subsubsection{Chalcopyrite}

\begin{figure*}
\centering
\includegraphics[width=1.5\columnwidth]{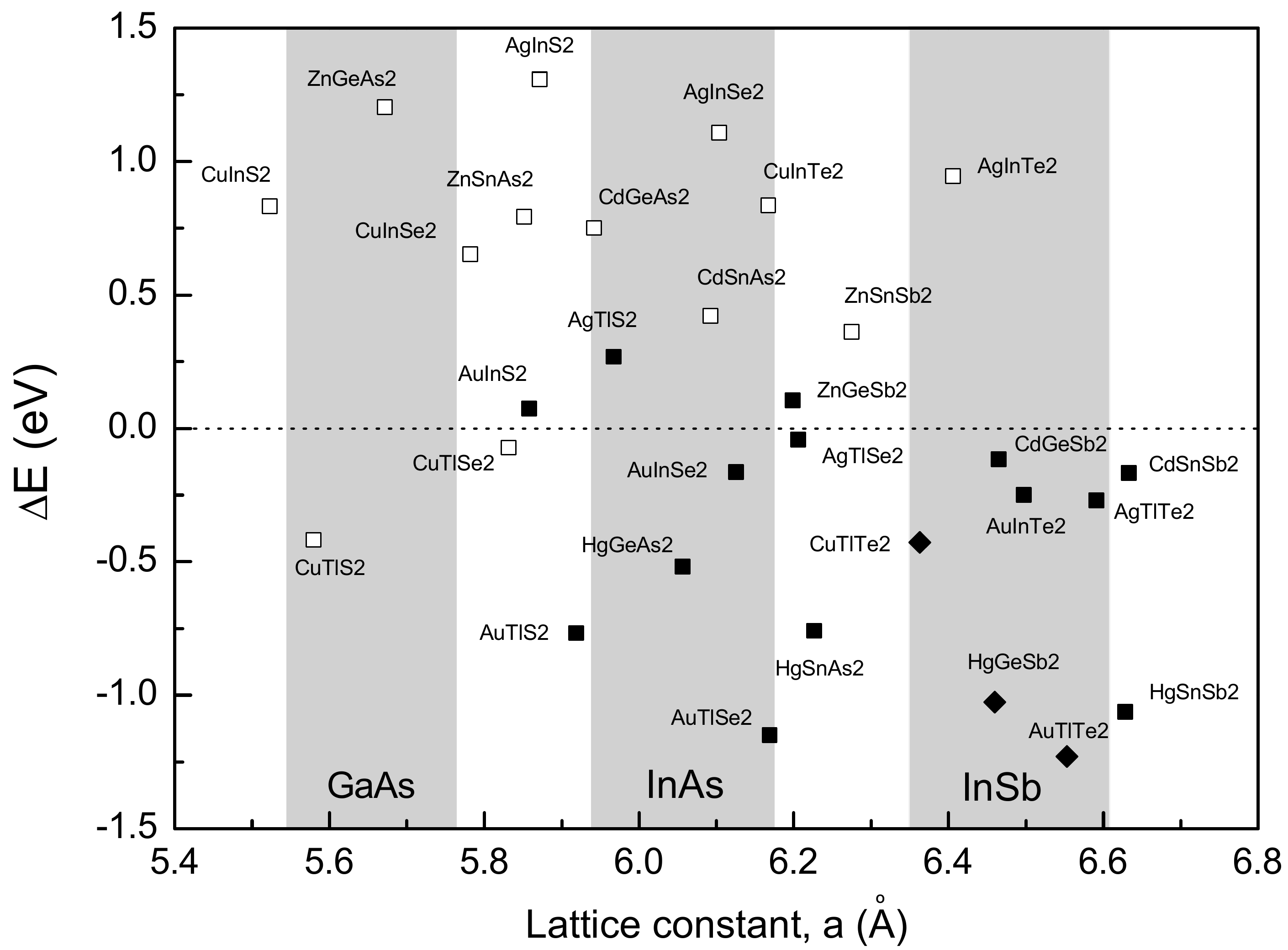}
\caption{Inverted band strength for various chalcopyrites as a function of the lattice constant. Open symbols mark the compounds whose lattice constant was reported in the literature. For the rest of the compounds their equilibrium lattice constants are obtained by first-principles total energy minimization. When $\Delta E<0$, squares mark the compounds that are topological insulators and diamonds mark the systems that are topological metals. Shaded areas indicate materials that are expected to be closely $(\pm2\%)$ lattice matched to either GaAs, InAs, or InSb. Adapted from Ref. [42].}\label{fig6}
\end{figure*}

The crystal structure of ternary chalcopyrite compounds is described by the space group $I\bar{4}2d$ (No. 122) with body-centered tetragonal structure.  The chemical formula of these materials is $ABC_{2}$, which can be regarded as a superlattice of two cubic zinc-blende unit cells and the A and B cations are ordered on the two different sites.  Since the overall structural similarity between the ternary chalcopyrites and binary zinc-blende analogs, the electronic structures of the former one are expected to closely resemble that of latter one.  Therefore, some of the chalcopyrite compounds may have the same topological class with the 3DTI HgTe.

Feng et al. [42] predicted that a large number of chalcopyrite compounds can realize the topological insulating phase by exploiting adiabatic continuity of band structures and direct evaluation of the $Z_{2}$ topological invariants.  Comparing with other 3DTI with cubic symmetry, a seeming advantage in chalcopyrite compounds is that topologically nontrivial state can be formed in their native states without any external strain.  This is because the AB cation ordering and additional structural modifications explicitly break the cubic symmetry.  In chalcopyrite compounds, the band inversion strength $\Delta E$ is redefined as the energy difference between the $s$-orbital originated $\Gamma_{6}$ states and the VBM at the $\Gamma$ point.  Figure \ref{fig6} shows the band topology of the chalcopyrite family of the I-III-VI$_{2}$ compounds (I = Cu, Ag, Au; III = In, Tl; V = S, Se, Te), as well as the II-IV-V$_{2}$ compounds (II = Zn, Cd, Hg; IV = Ge, Sn; V = As, Sb).  One can see that there are a large number of topologically nontrivial materials.  More importantly, many  chalcopyrite topological insulators have a close lattice matching to several mainstream semiconductors, which is essential for a smooth integration into current semiconductor technology.  The diverse physical properties of chalcopyrite semiconductors [108--111] make them appealing candidates for novel quantum devices.

\subsubsection{Thallium-based chalcogenides}

Thallium-based chalcogenides are narrow band-gap semiconductors with rhombohedral structure described by space group R$\bar{3}$m (No. 166).  The elemental composition of these materials is Tl$XY_{2}$ ($X$ = Bi, Sb and $Y$ = S, Se, Te), which can be regarded as a sequence of hexagonally close-packed layers with the order of -Tl-Y-X-Y-.  In contrast to Bi$_{2}$Se$_{3}$ family, the Tl$XY_{2}$ are not layered compounds because each Tl or X atomic layer is sandwiched by two Y atomic layers and every two atomic layers are coupled by strong covalent bonds.  As a consequence, the surface electronic structure is sensitive to surface relaxations and different terminations.

Although the identification of topologically nontrivial phase in these compounds is more complex than that of Bi$_{2}$Se$_{3}$ family, several first-principles studies have predicted that TlSbTe$_{2}$, TlSbSe$_{2}$, TlBiTe$_{2}$, and TlBiSe$_{2}$ are 3DTI [112--114].  The single Dirac cone surface state of TlBiTe$_{2}$, and TlBiSe$_{2}$ have been confirmed by ARPES measurements [115--118].

\subsection{3DTI in quaternary compounds\label{sec:3.5}}

The search for 3DTI has already been extended to quaternary I$_{2}$-II-IV-VI$_{4}$ compounds [133,134].  These materials are described by space group I$\bar{4}$2m (No. 121) with body-centered tetragonal structure.  Structurally, quaternary I$_{2}$-II-IV-VI$_{4}$ compounds can be viewed as the subcompounds derived from chalcopyrite I-III-VI$_{2}$ compounds by mutating two group-III cations to one group-II and one group-IV cations.  Because of the structural similarity, the band topology of these compounds is expected to be nontrivial in analogy to that of chalcopyrite 3DTI [42].  Based on first-principles calculations, Chen et al. [134] and Wang et al. [133] predicted that a number of I$_{2}$-II-IV-VI$_{4}$ compounds are indeed topologically nontrivial materials.  More importantly, the electronic properties of these compounds can be better altered because of the enhanced chemical and structural degrees of freedom.  For example, the nontrivial band-gaps can be further increased in Ag$_{2}$HgPbSe$_{4}$ (0.047 eV) and Cu$_{2}$ZnGeSe$_{4}$ (0.069 eV), which are larger than the band-gaps in strained HgTe and some of chalcopyrite compounds.

\end{multicols}

\begin{center} \footnotesize \doublerulesep 0.2pt \tabcolsep 12pt
\begin{longtable}[t]{ccccccc}
\caption{3DTI host materials in current literatures are classified by simple elements, binary compounds, pseudo-binary compounds, ternary compounds, and quaternary compounds.  Here, the tetradymite-like materials are called pseudo-binary compounds because their crystal and electronic structures are very similar to that of binary Bi$_2$Se$_3$ family.  In some materials with zero band-gaps, proper strains should be added to realize the topological insulating state.  Note that ternary half-Heusler and chalcopyrite 3DTI are not list one by one due to their large numbers, but the readers can refer to the Fig. \ref{fig5} and Fig. \ref{fig6}.} \label{tab1} \\

\toprule[0.9pt]

Coupounds & Bravais lattice & Space group (No.) & Band gap (eV) & Strain & Theory works & Experiment works  \\

\hline
$\alpha$-Sn$$ & fcc & Fd$\bar{3}$m (227) &      zero$^{[17]}$ & yes & [17]    &  --    \\
           Sb & rho &  R$\bar{3}$m (166) & semimetal$^{[17]}$ &  no & [17,37] & [51,52] \\

\hline
Bi$_{1-x}$Sb$_{x}$ & rho & -- & 0.03 ($x$=0.18)$^{[17]}$ & no & [17,26,27] & [52,55--57] \\
  Bi$_{2}$Se$_{3}$ & rho &  R$\bar{3}$m (166) & 0.30$^{[66]}$ &  no & [35,66] & [58]  \\
  Bi$_{2}$Te$_{3}$ & rho &  R$\bar{3}$m (166) & 0.12$^{[66]}$ &  no & [35,66] & [59,60]  \\
  Sb$_{2}$Te$_{3}$ & rho &  R$\bar{3}$m (166) & 0.17$^{[66]}$ &  no & [35,66] & [60]  \\
              HgTe & fcc & F$\bar{4}$3m (216) & zero$^{[17]}$  & yes & [17] & [72]  \\
       $\beta$-HgS & fcc & F$\bar{4}$3m (216) & 0.042$^{[73]}$  & no & [73] & --  \\
              InSb & fcc & F$\bar{4}$3m (216) & zero$^{[75]}$  & yes & [75] & --  \\
$\beta$-Ag$_{2}$Te & fcc &    P2$_{1}$/c (14) & 0.08$^{[76]}$  &  no & [76] & --  \\
        Sr$_{2}$Pb & orc &          Pnma (62) & 0.05$^{[77]}$  & yes & [77] & --  \\
       $\beta$-GaS & hex & P6$_{3}$/mmc (194) & 0.025$^{[78]}$ & yes & [78] & --  \\
   $\epsilon$-GaSe & hex & P$\bar{6}$m2 (187) & 0.135$^{[78]}$ & yes & [78] & --  \\
        Na$_{3}$Bi & hex & P6$_{3}$/mmc (194) & 0.006$^{[79]}$ & yes & [79] & -- \\
         K$_{3}$Bi & hex & P6$_{3}$/mmc (194) & -- & yes & [79] & -- \\
        Rb$_{3}$Bi & hex & P6$_{3}$/mmc (194) & -- & yes & [79] & -- \\

\hline
Bi$_2$Te$_2$S  & rho &  R$\bar{3}$m (166) & 0.28$^{[89]}$ &  no & [88,89] & -- \\
Bi$_2$Te$_2$Se & rho &  R$\bar{3}$m (166) & 0.28$^{[89]}$ &  no & [88--90] & [86,87,90,91,94] \\
Bi$_2$Se$_2$Te & rho &  R$\bar{3}$m (166) & 0.17$^{[89]}$ &  no & [89,90] & [90] \\
Sb$_2$Te$_2$S  & rho &  R$\bar{3}$m (166) & 0.29$^{[88]}$ &  no & [88] & -- \\
Sb$_2$Te$_2$Se & rho &  R$\bar{3}$m (166) & 0.34$^{[88]}$ &  no & [88] & -- \\
Bi$_2$Te$_{1.5}$S$_{1.5}$   & rho & -- & 0.2$^{[90]}$ &  no & [90] & [90] \\
Bi$_2$Te$_{1.6}$S$_{1.4}$   & rho & -- & 0.2$^{[91]}$ &  no &  --  & [91] \\
Bi$_{2-x}$Sb$_{x}$Te$_{3}$  & rho & -- & -- &  no & [93] & [92,93] \\
Bi$_{2-x}$Sb$_{x}$Te$_{3-y}$Se$_{y}$ & rho & -- & -- &  no & -- & [94--96] \\

\hline
half-Heusler (see Fig. \ref{fig5}) & fcc & F$\bar{4}$3m (216) & zero$^{[33,34]}$ & yes & [33,34,97--99] &  [100--104] \\
Li$_{2}$AgSb & fcc & F$\bar{4}$3m (216) & zero$^{[107]}$ & yes & [107] &  -- \\
chalphyrite (see Fig. \ref{fig6}) & bct & I$\bar{4}$2d (122) & 0.01$\sim$0.14$^{[42]}$ & no & [42] & -- \\
TlBiSe$_{2}$ & rho &  R$\bar{3}$m (166) & 0.2$^{[118]}$ &  no & [112--114] & [115--118] \\
TlBiTe$_{2}$ & rho &  R$\bar{3}$m (166) & semimetal$^{[118]}$ &  no & [112--114] & [118] \\
TlSbSe$_{2}$ & rho &  R$\bar{3}$m (166) & 0.14$^{[113]}$ &  no & [112--114] & -- \\
TlSbTe$_{2}$ & rho &  R$\bar{3}$m (166) & 0.05$^{[113]}$ &  no & [112--114] & -- \\
LaBiTe$_{3}$ & rho &  R$\bar{3}$m (166) & 0.12$^{[119]}$ &  no & [119] & -- \\
CeOs$_{4}$As$_{12}$ & bcc & Im$\bar{3}$ (204) &  zero$^{[120]}$ & yes & [120] & -- \\
CeOs$_{4}$Sb$_{12}$ & bcc & Im$\bar{3}$ (204) &  zero$^{[120]}$ & yes & [120] & -- \\
Ca$_{3}$NBi & cub &  Pm$\bar{3}$m (221) & 0.03$^{[121]}$ &  yes & [121] & -- \\
Sr$_{3}$NBi & cub &  Pm$\bar{3}$m (221) & -- &  yes & [121] & -- \\
Ba$_{3}$NBi & cub &  Pm$\bar{3}$m (221) & -- &  yes & [121] & -- \\
CsSnCl$_{3}$ & cub & Pm$\bar{3}$m (221) & 0.111$^{[122]}$ &  yes & [122] & -- \\
CsPbCl$_{3}$ & cub & Pm$\bar{3}$m (221) & 0.354$^{[122]}$ &  yes & [122] & -- \\
CsGeBr$_{3}$ & cub & Pm$\bar{3}$m (221) & 0.026$^{[122]}$ &  yes & [122] & -- \\
CsSnBr$_{3}$ & cub & Pm$\bar{3}$m (221) & 0.099$^{[122]}$ &  yes & [122] & -- \\
CsPbBr$_{3}$ & cub & Pm$\bar{3}$m (221) & 0.120$^{[122]}$ &  yes & [122] & -- \\
CsSnI$_{3}$ & cub & Pm$\bar{3}$m (221) & 0.169$^{[122]}$ &  yes & [122] & -- \\
LiAgSe & hex & P6$_{3}$/mmc (194) & 0.001$^{[123]}$ & yes & [123] &  --  \\
LiAuSe & hex & P6$_{3}$/mmc (194) & 0.050$^{[123]}$ & yes & [123] &  --  \\
LiAuTe & hex & P6$_{3}$/mmc (194) &  --             & yes & [123] &  --  \\
NaAgSe & hex & P6$_{3}$/mmc (194) & 0.010$^{[123]}$ & yes & [123] &  --  \\
NaAgTe & hex & P6$_{3}$/mmc (194) & 0.003$^{[123]}$ & yes & [123] &  --  \\
NaAuSe & hex & P6$_{3}$/mmc (194) & 0.015$^{[123]}$ & yes & [123] &  --  \\
NaAuTe & hex & P6$_{3}$/mmc (194) & 0.030$^{[123]}$ & yes & [123] &  --  \\
 KAuTe & hex & P6$_{3}$/mmc (194) &  --             & yes & [123] &  --  \\
LiHgAs & hex & P6$_{3}$/mmc (194) &  --             & yes & [123] &  --  \\
LiHgSb & hex & P6$_{3}$/mmc (194) &  --             & yes & [123] &  --  \\
BiTeI    & hex & P3m1 (156) & -- &  yes & [124] & -- \\
Ge$_{2}$Sb$_{2}$Te$_{5}$(Petrov sequence)  & hex & P$\bar{3}$m1 (164) & 0.1$^{[125]}$ &  no & [125] & -- \\
Ge$_{2}$Sb$_{2}$Te$_{5}$(KH sequence)  & hex & P$\bar{3}$m1 (164) & -- &  no & [126,127] & -- \\

PbBi$_{2}$Se$_{4}$ & rho & R$\bar{3}$m (166) & 0.40$^{[128]}$ &  no & [128,131,132] & [131,132] \\
PbBi$_{2}$Te$_{4}$ & rho & R$\bar{3}$m (166) & 0.23$^{[129]}$ &  no & [129,131,132] & [129--132] \\
Pb(Bi$_{1-x}$Sb$_{x}$)$_{2}$Te$_{4}$ & rho & -- & -- &  no & -- & [130] \\
GeBi$_{2}$Te$_{4}$ & rho & R$\bar{3}$m (166) & -- &  no & [131,132] & [131,132] \\
GeSb$_{2}$Te$_{4}$ & rho & R$\bar{3}$m (166) & -- &  no & [131,132] & [131,132] \\
SnBi$_{2}$Te$_{4}$ & rho & R$\bar{3}$m (166) & -- &  no & [131,132] & [131,132] \\
SnSb$_{2}$Te$_{4}$ & rho & R$\bar{3}$m (166) & -- &  no & [131,132] & [131,132] \\
PbSb$_{2}$Te$_{4}$ & rho & R$\bar{3}$m (166) & -- &  no & [131,132] & [131,132] \\
GeBi$_{4}$Te$_{7}$ & hex & P$\bar{3}$m1 (164) & -- &  no & [131,132] & [131,132] \\
GeSb$_{4}$Te$_{7}$ & hex & P$\bar{3}$m1 (164) & -- &  no & [131,132] & [131,132] \\
SnBi$_{4}$Te$_{7}$ & hex & P$\bar{3}$m1 (164) & -- &  no & [131,132] & [131,132] \\
SnSb$_{4}$Te$_{7}$ & hex & P$\bar{3}$m1 (164) & -- &  no & [131,132] & [131,132] \\
PbBi$_{4}$Se$_{7}$ & hex & P$\bar{3}$m1 (164) & -- &  no & [131,132] & [131,132] \\
PbBi$_{4}$Te$_{7}$ & hex & P$\bar{3}$m1 (164) & -- &  no & [131,132] & [131,132] \\
PbSb$_{4}$Te$_{7}$ & hex & P$\bar{3}$m1 (164) & -- &  no & [131,132] & [131,132] \\
GeBi$_{6}$Te$_{10}$ & rho & R$\bar{3}$m (166) & -- &  no & [131] & [131] \\
SnBi$_{6}$Te$_{10}$ & rho & R$\bar{3}$m (166) & -- &  no & [131] & [131] \\
PbBi$_{6}$Te$_{10}$ & rho & R$\bar{3}$m (166) & -- &  no & [131] & [131] \\
Cu$_{3}$SbS$_{4}$ & bct & I$\bar{4}$2m (121) & 0.042$^{[133]}$ & no & [133] & -- \\

\hline
Cu$_{2}$ZnGeSe$_{4}$ & bct & I$\bar{4}$2m (121) & 0.069$^{[133]}$ & no & [133] & -- \\
Cu$_{2}$ZnSnSe$_{4}$ & bct & I$\bar{4}$2m (121) & 0.056$^{[133]}$ & no & [133] & -- \\
Cu$_{2}$CdGeSe$_{4}$ & bct & I$\bar{4}$2m (121) & 0.003$^{[133]}$ & no & [133] & -- \\
Cu$_{2}$CdSnSe$_{4}$ & bct & I$\bar{4}$2m (121) & 0.003$^{[133]}$ & no & [133] & -- \\
Cu$_{2}$HgGeSe$_{4}$ & bct & I$\bar{4}$2m (121) & 0.006$^{[133]}$ & no & [133] & -- \\
Cu$_{2}$HgSnSe$_{4}$ & bct & I$\bar{4}$2m (121) & 0.007$^{[133]}$ & no & [133] & -- \\
Cu$_{2}$CdSnS$_{4}$ & bct & I$\bar{4}$2m (121) & 0.033$^{[133]}$ & no & [133] & -- \\
Cu$_{2}$HgSnS$_{4}$ & bct & I$\bar{4}$2m (121) & 0.060$^{[133]}$ & no & [133] & -- \\
Ag$_{2}$HgPbSe$_{4}$ & bct & I$\bar{4}$2m (121) & 0.047$^{[134]}$ & no & [134] & -- \\
Ag$_{2}$CdPbTe$_{4}$ & bct & I$\bar{4}$2m (121) & -- & no & [134] & -- \\
Cu$_{2}$HgPbSe$_{4}$ & bct & I$\bar{4}$2m (121) & -- & no & [134] & -- \\
Cu$_{2}$CdPbSe$_{4}$ & bct & I$\bar{4}$2m (121) & -- & no & [134] & -- \\
\bottomrule[0.9pt] 
\end{longtable}
\end{center}

\begin{multicols}{2}

\section{Conclusion and outlook\label{sec:4}}

Since the potential applications in spintronic field and quantum computation, 3DTI has no doubt become topic of interest in the fields of condensed matter physics and material science.  Thousands of scientific colleagues are attracted by its exotic quantum phenomena, pushing this field in a rapid pace.  Similar to other stages of development in the history of condensed matter physics, for example the QH effect, the emergence of high quality samples is of crucial importance.  For this purpose, this paper reviews the recent progress of materials realization in 3DTI.

Based on topological band theory [18,19], the first-principles calculation plays a pivotal role in the prediction of 3DTI host materials.  We have introduced four methods of searching for new 3DTI, including surface state electronic structure, adiabatic continuity band transformation, band inversion picture, and $Z_2$ topological invariants.  None of these methods is solely used in present literatures, but two or more of them are usually combined together to predict topologically nontrivial materials.  To date, many materials have been predicted to be 3DTI (see Table \ref{tab1}).  Some of them have been confirmed by experiments via the ARPES measurements of surface Dirac electronic structure.  The most favorable materials are thus the Bi$_{2}$Se$_{3}$ family, which have been extensively studied from bulk to thinfilm [58--71].  But remarkably conducting bulk state even in high quality sample blocks further studies of the surface transport.  For decreasing the bulk carrier concentration, the pseudo-binary compound derived from Bi$_{2}$Se$_{3}$ family, for example Bi$_{2}$Te$_{2}$Se, is found to be new class of 3DTI with much larger bulk resistivity [86--94].  The theoretical prediction of 3DTI was also extended to ternary and quaternary compounds, but there are rare experimental reports about these compounds mainly because of the difficulties in their synthesis and growth.

We propose several interesting and valuable directions in the further work.  (i) Among the present known 3DTI host materials, there does not exist a unique one which can sufficiently meet various requirements in experimental observations.  Therefore, searching for new 3DTI host materials, particularly for those with simpler surface electronic structure and easier synthetic and growing condition, is still needed.  This should be a central task in this rapidly developing field.  (ii) Although the present 3DTI host materials are often discussed in the context of the non-interacting band theory, the materials realization of topological Mott insulator [135,136] and topological Kondo insulator [137,138] with strong electron-electron interaction should also be taken into account.  Furthermore, the topological order affected by disorder, i.e., topological Anderson insulator [139--141], is interesting and needs to be thoroughly studied.  (iii) The realization of the quantum anomalous Hall effect and topological superconductor is also a frontier direction.  For quantum anomalous Hall effect, some theoretical models were proposed by means of magnetic doping in the film or surface of 3DTI [142--147], but they have not been reproduced by current experiments.  For topological superconductor, one expects to realize it by constructing proper interface between 3DTI and conventional superconductors due to the proximity effect [18,148].  In addition, a mysterious particle obeyed non-Abelian statistics, the Majorana fermion, could be detected in the interface and it will be hopefully utilized in the topological quantum computations.

\Acknowledgements{\bahao The authors were supported by National Basic Research Program of China (973 Program Grants No. 2011CBA00100) and National Natural Science Foundation of China (Grants No. 10974231 and 11174337).}


\normalsize \vskip0.1in\parskip=0mm \baselineskip 18pt
\renewcommand{\baselinestretch}{1.06}\footnotesize\parindent=4mm\bahao

\vskip0.1in \noindent 
\vskip0.1in\parskip=0mm

\REF{1\ }Wen X G. Topological orders and edge excitations in fractional quantum Hall states. Adv Phys, 1995, 44: 405--473
\REF{2\ }Thouless D J, Kohmoto M, Nightingale M P, Nijs M d. Quantized hall conductance in a two-dimensional periodic potential. Phys Rev Lett, 1982, 49: 405--408
\REF{3\ }Klitzing K v, Dorda G, Pepper M. New method for high-accuracy determination of the fine-structure constant based on quantized Hall resistance. Phys Rev Lett, 1980, 45: 494--497
\REF{4\ }Bernevig B A, Zhang S C. Quantum spin Hall effect. Phys Rev Lett, 2006, 96: 106802
\REF{5\ }Kane C L, Mele E J. A new spin on the insulating state. Science, 2006, 314: 1692--1693
\REF{6\ }Kane C L, Mele E J. Quantum spin Hall effect in graphene. Phys Rev Lett, 2005, 95: 226801
\REF{7\ }Kane C L, Mele E J. Z$_2$ topological order and the quantum spin Hall effect. Phys Rev Lett, 2005, 95: 146802
\REF{8\ }Kohn W. Theory of the insulating state. Phys Rev, 1964, 133: A171--A181
\REF{9\ } Yao Y G, Ye F, Qi X L, et al. Spin-orbit gap of graphene: First-principles calculations. Phys Rev B, 2007, 75: 041401
\REF{10\ } Huertas-Hernando D, Guinea F, Brataas A. Spin-orbit coupling in curved graphene, fullerenes, nanotubes, and nanotube caps. Phys Rev B, 2006, 74: 155426
\REF{11\ } Min H, Hill J E, Sinitsyn N A, et al. Intrinsic and Rashba spin-orbit interactions in graphene sheets. Phys Rev B, 2006, 74: 165310
\REF{12\ }Bernevig B A, Hughes T L, Zhang S C. Quantum spin Hall effect and topological phase transition in HgTe quantum wells. Science, 2006, 314: 1757--1761
\REF{13\ }K\"{o}nig M, Wiedmann S, Br\"{u}ne C, et al. Quantum spin Hall insulator state in HgTe quantum wells. Science, 2006, 318: 766-770
\REF{14\ }Fu L, Kane C L, Mele E J. Topological insulators in three dimensions. Phys Rev Lett, 2007, 98: 106803
\REF{15\ }Moore J E, Balents L. Topological invariants of time-reversal-invariant band structures. Phys Rev B, 2007, 75: 121306
\REF{16\ }Roy R. Topological phases and the quantum spin Hall effect in three dimensions. Phys Rev B, 2009, 79: 195322
\REF{17\ }Fu L, Kane C L. Topological insulators with inversion symmetry. Phys Rev B, 2007, 76: 045302
\REF{18\ }Qi X L, Zhang S C. Topological insulators and superconductors. Rev Mod Phys, 2011, 83: 1057--1110
\REF{19\ }Qi X L, Zhang S C. The quantum spin Hall effect and topological insulators. Physics Today, 2010, 63: 33--38
\REF{20\ }Hasan M Z, Kane C L. Colloquium : Topological insulators. Rev Mod Phys, 2010, 82: 3045--3067
\REF{21\ }Hasan M Z, Moore J E. Three-dimensional topological insulators. Ann Rev Cond Matter Phys, 2011, 2: 55--78
\REF{22\ }Moore J E. The birth of topological insulators. Nature (London), 2010, 464: 194--198
\REF{23\ }Brumfiel G, Topological insulators: Star material. Nature (London), 2010, 466: 310--311
\REF{24\ }DiVincenzo D P, Mele E J. Self-consistent effective-mass theory for intralayer screening in graphite intercalation compounds. Phys Rev B, 1984, 29: 1685--1694
\REF{25\ }Semenoff G W. Condensed-matter simulation of a three-dimensional anomaly. Phys Rev Lett, 1984, 53: 2449--2452
\REF{26\ }Appelbaum Joel A, Hamann D R. The electronic structure of solid surfaces. Rev Mod Phys, 1976, 48: 479--496
\REF{27\ }Hoffmann R. A chemical and theoretical way to look at bonding on surfaces. Rev Mod Phys, 1988, 60: 601--628
\REF{28\ }Hoffmann R. Solids and surfaces: a chemist's view of bonding in extended structures. New York: VCH Publishers, 1988.
\REF{29\ }Liu C C, Feng W X, Yao Y G. Quantum spin Hall effect in silicene and two-dimensional germanium. Phys Rev Lett, 2011, 107: 076802
\REF{30\ }Teo J C Y, Fu L, Kane C L. Surface states and topological invariants in three-dimensional topological insulators: Application to Bi$_{1-x}$Sb$_{x}$. Phys Rev B, 2008, 78: 045426
\REF{31\ }Zhang H J, Liu C X, Qi X L, et al. Electronic structures and surface states of the topological insulator Bi$_{1-x}$Sb$_{x}$. Phys Rev B, 2009, 80: 085307
\REF{32\ }Delin A, Kl\"{u}ner T. Excitation spectra and ground-state properties from density-functional theory for the inverted band-structure systems $\beta$-HgS, HgSe, and HgTe. Phys Rev B, 2002, 66: 035117.
\REF{33\ }Xiao D, Yao Y G, Feng W X, et al. Half-Heusler compounds as a new class of three-dimensional topological insulators. Phys Rev Lett, 2010, 105: 096404
\REF{34\ }Feng W X, Xiao D, Zhang Y, et al. Half-Heusler topological insulators: A first-principles study with the Tran-Blaha modified Becke-Johnson density functional. Phys Rev B, 2010, 82: 235121
\REF{35\ }Zhang H J, Liu C X, Qi X L, et al. Topological insulators in Bi$_2$Se$_3$, Bi$_2$Te$_3$ and Sb$_2$Te$_3$ with a single Dirac cone on the surface. Nature Phys, 2009, 5: 438--442
\REF{36\ } Zhang X, Zhang H J, Wang J, et al. Actinide topological insulator materials with strong interaction. Science, 2012, 335: 1464--1466
\REF{37\ }Fukui T, Hatsugai Y. Quantum spin Hall effect in three dimensional materials: Lattice computation of Z$_2$ topological invariants and its application to Bi and Sb. J. Phys. Soc. Japan, 2007, 76: 053702
\REF{38\ }Feng W X, Wen J, Zhou J J, et al. First-principles calculation of Z$_2$ topological invariants within the FP-LAPW formalism. Comp Phys Comm, 2012, 183: 1849--1859
\REF{39\ }Fu L, Kane C L. Time reversal polarization and a Z$_2$ adiabatic spin pump. Phys. Rev. B, 2006, 74: 195312
\REF{40\ }King-Smith R D, Vanderbilt D. Theory of polarization of crystalline solids. Phys Rev B, 1993, 47: 1651--1654
\REF{41\ }Resta R. Macroscopic polarization in crystalline dielectrics: the gemetric phase approach. Rev Mod Phys, 1994, 66: 899--915
\REF{42\ }Feng W X, Xiao D, Ding J, et al. Three-dimensional topological insulators in I-III-VI$_2$ and II-IV-V$_2$ chalcopyrite semiconductors. Phys Rev Lett, 2011, 106: 016402
\REF{43\ }Soluyanov A A, Vanderbilt D. Computing topological invariants without inversion symmetry. Phys Rev B, 2011,83: 235401
\REF{44\ }Yu R, Qi X L, Bernevig A, et al. Equivalent expression of Z$_2$ topological invariant for band insulators using the non-Abelian Berry connection. Phys Rev B, 2011, 84: 075119
\REF{45\ }Singh D J. Planewaves, Pseudopotentials and the LAPW Method. Boston: Kluwer Academic, 1994.
\REF{46\ }Tran F, Blaha P. Accurate band gaps of semiconductors and insulators with a semilocal exchange-correlation potential. Phys Rev Lett, 2009, 102: 226401
\REF{47\ }Blaha P, Schwarz K, Madsen G, et al. Wien2k, an augmented plane wave plus local orbitals program for calculating crystal properties. Austria: Vienna University of Technology, Vienna, 2001.
\REF{48\ }Groves S, Paul W. Band structure of gray tin. Phys Rev Lett, 1963, 11: 194--196
\REF{49\ }Chelikowsky J R, Cohen M L. Nonlocal pseudopotential calculations for the electronic structure of eleven diamond and zinc-blende semiconductors. Phys Rev B, 1976, 14: 556--582
\REF{50\ }Perdew J P, Burke K, Ernzerhof M. Generalized gradient approximation made simple. Phys Rev Lett, 1996, 77: 3865--3868
\REF{51\ }Hsieh D, Wray L, Qian D, et al. Direct observation of spin-polarized surface states in the parent compound of a topological insulator using spin- and angle-resolved photoemission spectroscopy in a Mott-polarimetry mode. New J Phys, 2010, 12: 125001
\REF{52\ }Hsieh D, Xia Y, Wray L, et al. Observation of unconventional quantum spin textures in topological insulators. Science, 2009, 323: 919--922
\REF{53\ }Lerner L S, Cuff K F, Williams L R. Energy-band parameters and relative band-edge motions in the Bi-Sb alloy system near the semimetal―semiconductor transition. Rev Mod Phys, 1968, 40: 770--775
\REF{54\ }Liu Y, Allen R E. Electronic structure of the semimetals Bi and Sb. Phys Rev B, 1995, 52: 1566--1577
\REF{55\ }Hsieh D, Qian D, Wray L, et al. A topological Dirac insulator in a quantum spin Hall phase. Nature (London) 2008, 452: 970--974
\REF{56\ }Guo H, Sugawara K, Takayama A, et al. Evolution of surface states in Bi$_{1-x}$Sb$_{x}$ alloys across the topological phase transition. Phys Rev B, 2011, 83: 201104
\REF{57\ }Nishide A, Taskin A A, Takeichi Y, et al. Direct mapping of the spin-filtered surface bands of a three-dimensional quantum spin Hall insulator. Phys Rev B, 2010, 81: 041309
\REF{58\ }Xia Y, Qian D, Hsieh D, et al. Observation of a large-gap topological-insulator class with a single Dirac cone on the surface. Nature Phys. 2009, 5: 398--402
\REF{59\ }Chen Y L, Analytis J G, Chu J H, et al. Experimental realization of a three-dimensional topological insulator, Bi$_2$Te$_3$. Science 2009, 325: 178--180
\REF{60\ }Hsieh D, Xia Y, Qian D, et al. Observation of time-reversal-protected single-Dirac-cone topological-insulator states in Bi$_2$Te$_3$ and Sb$_2$Te$_3$. Phys Rev Lett, 2009, 103: 146401
\REF{61\ }Zhang T, Cheng P, Chen X, et al. Experimental demonstration of topological surface states protected by time-reversal symmetry. Phys Rev Lett, 2009, 103: 266803
\REF{62\ }Chen Y L, Chu J H, Analytis J G, et al. Massive Dirac fermion on the surface of a magnetically doped topological insulator. Science 2010, 329: 659--662
\REF{63\ }Hsieh D, Xia Y, Qian D, et al. A tunable topological insulator in the spin helical Dirac transport regime. Nature (London) 2009, 460: 1101--1105
\REF{64\ }Kuroda K, Arita M, Miyamoto K, et al. Hexagonally deformed Fermi surface of the 3D topological insulator Bi$_2$Se$_3$. Phys Rev Lett, 2010, 105: 076802
\REF{65\ }Zhang J L, Zhang S J, Weng H M, et al. Pressure-induced superconductivity in topological parent compound Bi$_2$Te$_3$. PNAS, 2011, 108: 24--28
\REF{66\ }Zhang W, Yu R, Zhang H J, et al. First-principles studies of the three-dimensional strong topological insulators Bi$_2$Te$_3$, Bi$_2$Se$_3$ and Sb$_2$Te$_3$. New J Phys, 2010, 12: 065013
\REF{67\ }Park K, Heremans J J, Scarola V W, et al. Robustness of topologically protected surface states in layering of Bi$_2$Te$_3$ thin films. Phys Rev Lett, 2010, 105: 186801
\REF{68\ }Yu R, Zhang W, Zhang H J, et al. Quantized anomalous Hall effect in magnetic topological insulators. Science, 2010, 329: 61--64
\REF{69\ }Liu C X, Qi X L, Zhang H J, et al. Model Hamiltonian for topological insulators. Phys Rev B, 2010, 82: 045122
\REF{70\ }Liu C X, Zhang H J, Yan B H, et al. Oscillatory crossover from two-dimensional to three-dimensional topological insulators. Phys Rev B, 2010, 81: 041307
\REF{71\ }Song J H, Jin H, Freeman A J. Interfacial Dirac cones from alternating topological invariant superlattice structures of Bi$_2$Se$_3$. Phys Rev Lett, 2010, 105: 096403
\REF{72\ }Br\"{u}ne C, Liu C X, Novik E G, et al. Quantum Hall effect from the topological surface states of strained bulk HgTe. Phys Rev Lett, 2011, 106: 126803
\REF{73\ }Virot F, Hayn R, Richter M, et al. Metacinnabar ($\beta$-HgS): A strong 3D topological insulator with highly anisotropic surface states. Phys Rev Lett, 2011, 106: 236806
\REF{74\ }Madelung O. Semiconductors: Data handbook. Berlin: Springer, 2004.
\REF{75\ }Feng W X, Zhu W G, Weitering H H, et al. Strain tuning of topological band order in cubic semiconductors. Phys Rev B, 2012, 85: 195114
\REF{76\ }Zhang W, Yu R, Feng W X, et al. Topological aspect and quantum magnetoresistance of $\beta$-Ag$_2$Te. Phys Rev Lett, 2011, 106: 156808
\REF{77\ }Sun Y, Chen X Q, Franchini C, et al. Strain-driven onset of nontrivial topological insulating states in Zintl Sr$_2$X compounds (X = Pb, Sn). Phys Rev B, 2011, 84: 165127
\REF{78\ }Zhu Z Y, Cheng Y C, Schwingenschl\"{o}gl U. Topological phase transition in layered GaS and GaSe. Phys Rev Lett, 2012, 108: 266805
\REF{79\ }Wang Z, Sun Y, Chen X Q, et al. Dirac semimetal and topological phase transitions in A$_{3}$Bi (A = Na, K, Rb). Phys Rev B, 2012, 85: 195320
\REF{80\ }Hor Y S, Richardella A, Roushan P, et al. p-type Bi$_2$Se$_3$ for topological insulator and low-temperature thermoelectric applications. Phys Rev B, 2009, 79: 195208
\REF{81\ }Checkelsky J G, Hor Y S, Liu M H, et al. Quantum interference in macroscopic crystals of nonmetallic Bi$_2$Se$_3$. Phys Rev Lett, 2009, 103: 246601
\REF{82\ }Analytis J G, Chu J H, Chen Y L, et al. Bulk Fermi surface coexistence with Dirac surface state in Bi$_2$Se$_3$: A comparison of photoemission and Shubnikov-de Haas measurements. Phys Rev B, 2010, 81: 205407
\REF{83\ }Chen J, Qin H J, Yang F, et al. Gate-voltage control of chemical potential and weak antilocalization in Bi$_2$Se$_3$. Phys Rev Lett, 2010, 105: 176602
\REF{84\ }Chen J, He X Y, Wu K H, et al. Tunable surface conductivity in Bi$_2$Se$_3$ revealed in diffusive electron transport. Phys Rev B, 2011, 83: 241304
\REF{85\ }Checkelsky J G, Hor Y S, Cava R J, et al. Bulk band gap and surface state conduction observed in voltage-tuned crystals of the topological insulator Bi$_2$Se$_3$. Phys Rev Lett, 2011, 106: 196801
\REF{86\ }Ren Z, Taskin A A, Sasaki S, et al. Large bulk resistivity and surface quantum oscillations in the topological insulator Bi$_2$Te$_2$Se. Phys Rev B, 2010, 82: 241306
\REF{87\ }Xiong J, Petersen A C, Qu D X, et al. Quantum oscillations in a topological insulator Bi$_{2}$Te$_{2}$Se with large bulk resistivity (6 $\Omega$ cm). Physica E, 2012, 44: 917--920
\REF{88\ }Lin H, Das T, Wray L A, et al. An isolated Dirac cone on the surface of ternary tetradymite-like topological insulators. New J Phys, 2011, 13: 095005
\REF{89\ }Wang L L, Johnson D D. Ternary tetradymite compounds as topological insulators. Phys Rev B, 2011, 83: 241309
\REF{90\ }Neupane M, Xu S Y, Wray L A, et al. Topological surface states and Dirac point tuning in ternary topological insulators. Phys Rev B, 2012, 85: 235406
\REF{91\ }Ji H W, Allred J M, Fuccillo M K, et al. Bi$_{2}$Te$_{1.6}$S$_{1.4}$: A topological insulator in the tetradymite family. Phys Rev B, 2012, 85: 201103
\REF{92\ }Zhang J S, Chang C Z, Zhang Z C, et al. Band structure engineering in (Bi$_{1-x}$Sb$_{x}$)$_2$Te$_3$ ternary topological insulators. Nature Commun, 2011, 2: 574
\REF{93\ }Kong D S, Chen Y L, Cha J J, et al. Ambipolar field effect in the ternary topological insulator (Bi$_{x}$Sb$_{1-x}$)$_{2}$Te$_{3}$ by composition tuning.  Nature Nanotech, 2011, 6: 705--709
\REF{94\ }Arakane T, Sato T, Souma S, et al. Tunable Dirac cone in the topological insulator Bi$_{2-x}$Sb$_{x}$Te$_{3-y}$Se$_{y}$. Nature Commun, 2012, 3: 636
\REF{95\ }Ren Z, Taskin A A, Sasaki S, et al. Optimizing Bi$_{2-x}$Sb$_{x}$Te$_{3-y}$Se$_{y}$ solid solutions to approach the intrinsic topological insulator regime. Phys Rev B, 2011, 84: 165311
\REF{96\ }Taskin A A, Ren Z, Sasaki S, et al. Observation of Dirac holes and electrons in a topological insulator. Phys Rev Lett, 2011, 107: 016801
\REF{97\ }Chadov S, Qi X L, K\"{u}bler J, et al. Tunable multifunctional topological insulators in ternary Heusler compounds. Nature Mater, 2010, 9: 541--545
\REF{98\ }Lin H, Wray L A, Xia Y, et al. Half-Heusler ternary compounds as new multifunctional experimental platforms for topological quantum phenomena. Nature Mater. 2010, 9: 546--549
\REF{99\ }Al-Sawai W, Lin H, Markiewicz R S, et al. Topological electronic structure in half-Heusler topological insulators. Phys Rev B, 2010, 82: 125208
\REF{100\ }Gofryk K, Kaczorowski D, Plackowski T, et al. Magnetic and transport properties of rare-earth-based half-Heusler phases RPdBi: Prospective systems for topological quantum phenomena. Phys Rev B, 2011, 84: 035208
\REF{101\ }Liu C, Lee Y, Kondo T, et al. Metallic surface electronic state in half-Heusler compounds RPtBi (R = Lu, Dy, Gd). Phys Rev B, 2011, 83: 205133
\REF{102\ }Butch N P, Syers P, Kirshenbaum K, et al. Superconductivity in the topological semimetal YPtBi. Phys Rev B, 2011 84: 220504
\REF{103\ }Shekhar C, Ouardi S, Fecher G H, et al. Electronic structure and linear magnetoresistance of the gapless topological insulator PtLuSb. Appl Phys Lett, 2012, 100: 252109
\REF{104\ }Miyawaki T, Sugimoto N, Fukatani N, et al. Structural and electrical properties of half-Heusler La-Pt-Bi thin films grown by 3-source magnetron co-sputtering. J Appl Phys, 2012, 111: 07E327
\REF{105\ }Canfield P C, Thompson J D, Beyermann W P, et al. Magnetism and heavy fermion-like behavior in the RBiPt series. J Appl Phys, 1991, 70: 5800--5802
\REF{106\ }Goll G, Marz M, Hamann A, et al. Thermodynamic and transport properties of the non-centrosymmetric superconductor LaBiPt. Physica B, 2008, 403: 1065--1067
\REF{107\ }Lin H, Wray L A, Xia Y, et al. Single-Dirac-cone Z$_2$ topological insulator phases in distorted Li$_2$AgSb-class and related quantum critical Li-based spin-orbit compounds. E-print at arXiv:1004.0999
\REF{108\ }Shay J L, Wernick J H. Ternary chalcopyrite semiconductors: Growth, electronic properties and applications. Oxford: Pergamon Press, 1975.
\REF{109\ }Medvedkin G A, Ishibashi T, Nishi T, et al. Room temperature ferromagnetism in novel diluted magnetic semiconductor (Cd$_{1-x}$Mn$_x$)GeP$_2$. Jpn J Appl Phys, 2000, 39: L949--L951
\REF{110\ }Cho S, Choi S, Cha G B, et al. Room-temperature ferromagnetism in (Zn$_{1-x}$Mn$_x$)GeP$_2$ semiconductors. Phys Rev Lett, 2002, 88: 257203
\REF{111\ }Erwin S C, \v{Z}uti\'{c} I. Tailoring ferromagnetic chalcopyrites. Nature Mater, 2004, 3: 410--414
\REF{112\ }Lin H, Markiewicz R S, Wray L A, et al. Single-Dirac-cone topological surface states in the TlBiSe$_2$ class of topological semiconductors. Phys Rev Lett, 2010, 105: 036404
\REF{113\ }Yan B H, Liu C X, Zhang H J, et al. Theoretical prediction of topological insulators in thallium-based III-V-VI$_2$ ternary chalcogenides. Europhys Lett, 2010, 90: 37002
\REF{114\ }Eremeev S V, Bihlmayer G, Vergniory M, et al. Ab initio electronic structure of thallium-based topological insulators. Phys Rev B, 2011, 83: 205129
\REF{115\ }Kuroda K, Ye M, Kimura A, et al. Experimental realization of a three-dimensional topological insulator phase in ternary chalcogenide TlBiSe$_2$. Phys Rev Lett, 2010, 105: 146801
\REF{116\ }Sato T, Segawa K, Guo H, et al. Direct evidence for the Dirac-cone topological surface states in the ternary chalcogenide TlBiSe$_2$. Phys Rev Lett, 2010, 105: 136802
\REF{117\ }Xu S Y, Xia Y, Wray L A, et al. Topological phase transition and texture inversion in a tunable topological insulator. Science, 2011, 332: 560--564 
\REF{118\ }Chen Y L, Liu Z K, Analytis J G, et al. Single Dirac cone topological surface state and unusual thermoelectric property of compounds from a new topological insulator family. Phys Rev Lett, 2010, 105: 266401
\REF{119\ }Yan B H, Zhang H J, Liu C X, et al. Theoretical prediction of topological insulator in ternary rare earth chalcogenides. Phys Rev B, 2010, 82: 161108
\REF{120\ }Yan B H, M\"{u}chler L, Qi X L, et al. Topological insulators in filled skutterudites. Phys Rev B, 2012, 85: 165125
\REF{121\ }Sun Y, Chen X Q, Yunoki S, et al. New family of three-dimensional topological insulators with antiperovskite structure. Phys Rev Lett, 2010, 105: 216406
\REF{122\ }Yang K, Setyawan W, Wang S D, et al. A search model for topological insulators with high-throughput robustness descriptors. Nature Mater, 2012, 11: 614--619
\REF{123\ }Zhang H J, Chadov S, M\"{u}chler L, et al. Topological insulators in ternary compounds with a honeycomb lattice. Phys Rev Lett, 2011, 106: 156402
\REF{124\ }Bahramy M S, Yang B J, Arita R, et al. Emergence of non-centrosymmetric topological insulating phase in BiTeI under pressure. Nature Commun, 2012, 3: 679
\REF{125\ }Kim J, Kim J, Jhi S H. Prediction of topological insulating behavior in crystalline Ge-Sb-Te.  Phys Rev B, 2010, 82: 201312
\REF{126\ }Sa B S, Zhou J, Song Z T, et al. Pressure-induced topological insulating behavior in the ternary chalcogenide Ge$_2$Sb$_2$Te$_5$. Phys Rev B, 2011, 84: 085130
\REF{127\ }Sa B, Zhou J, Sun Z, et al. Strain-induced topological insulating behavior in ternary chalcogenide Ge$_2$Sb$_2$Te$_5$. Europhys Lett, 2011, 97: 27003
\REF{128\ }Jin H, Song J H, Freeman A J, et al. Candidates for topological insulators: Pb-based chalcogenide series. Phys Rev B, 2011, 83: 041202
\REF{129\ }Kuroda K, Miyahara H, Ye M, et al. Experimental verification of PbBi$_2$Te$_4$ as a 3D topological insulator. Phys Rev Lett, 2012, 108: 206803
\REF{130\ }Souma S, Eto K, Nomura M, et al. Topological surface states in lead-based ternary telluride Pb(Bi$_{1-x}$Sb$_{x}$)$_{2}$Te$_{4}$. Phys Rev Lett, 2012, 108: 116801
\REF{131\ }Eremeev S V, Landolt G, Menshchikova T V, et al. Atom-specific spin mapping and buried topological states in a homologous series of topological insulators. Nature Commun, 2012, 3: 635
\REF{132\ }Xu S Y, Wray L A, Xia Y, et al. Discovery of several large families of topological insulator classes with backscattering-suppressed spin-polarized single-dirac-cone on the surface. E-print at arXiv:1007.5111
\REF{133\ }Wang Y J, Lin H, Das T, et al. Topological insulators in the quaternary chalcogenide compounds and ternary famatinite compounds. New J Phys, 2011, 13: 085017
\REF{134\ }Chen S Y, Gong X G, Duan C G, et al. Band structure engineering of multinary chalcogenide topological insulators. Phys Rev B, 2011, 83: 245202
\REF{135\ }Raghu S, Qi X L, Honerkamp C, et al. Topological Mott insulators. Phys Rev Lett, 2008, 100: 156401
\REF{136\ }Zhang Y, Ran Y, Vishwanath A. Topological insulators in three dimensions from spontaneous symmetry breaking. Phys Rev B, 2009, 79: 245331
\REF{137\ }Dzero M, Sun K, Galitski V, et al. Topological Kondo insulators. Phys Rev Lett, 2010, 104: 106408
\REF{138\ }Dzero M, Sun K, Coleman P, et al. Theory of topological Kondo insulators. Phys Rev B, 2012, 85: 045130
\REF{139\ }Li J, Chu R L, Jain J K, et al. Topological Anderson insulator. Phys Rev Lett, 2009, 102: 136806
\REF{140\ }Groth C W, Wimmer M, Akhmerov A R, et al. Theory of the topological Anderson insulator. Phys Rev Lett, 2009, 103: 196805
\REF{141\ }Guo H M, Rosenberg G, Refael G, et al. Topological Anderson insulator in three dimensions. Phys Rev Lett, 2010, 105: 216601
\REF{142\ }Haldane F D M. Model for a quantum Hall effect without landau levels: Condensed-matter realization of the \textquotedblleft parity anomaly\textquotedblright. Phys Rev Lett, 1988, 61: 2015--2018
\REF{143\ }Liu C X, Qi X L, Dai X, et al. Quantum anomalous Hall effect in Hg$_{1-y}$Mn$_{y}$Te quantum wells. Phys Rev Lett, 2008, 101: 146802
\REF{144\ }Yu R, Zhang W, Zhang H J, et al. Quantized anomalous Hall effect in magnetic topological insulators. Science, 2010, 329: 61--64
\REF{145\ }Qiao Z H, Yang S Y A, Feng W X, et al. Quantum anomalous Hall effect in graphene from Rashba and exchange effects. Phys Rev B, 2010, 82: 161414
\REF{146\ }Ding J, Qiao Z H, Feng W X, et al. Engineering quantum anomalous/valley Hall states in graphene via metal-atom adsorption: An ab-initio study. Phys Rev B, 2011, 84: 195444
\REF{147\ }Zhang H J, Zhang X, Zhang S C. Quantum Anomalous Hall Effect in Magnetic Topological Insulator GdBiTe$_{3}$. E-print at arXiv:1108.4857
\REF{148\ }Fu L, Kane C L. Superconducting proximity effect and Majorana fermions at the surface of a topological insulator. Phys Rev Lett, 2008, 100: 096407

\end{multicols}

\end{document}